\documentclass[prd,showpacs,floatfix,twocolumn,amsmath,amssymb,floatfix]{revtex4}
\usepackage{amsmath}
\usepackage{graphicx,color,dcolumn,booktabs,bm}
\usepackage{longtable,lscape}
\usepackage{txfonts}
\usepackage{overpic}
\usepackage{amssymb}
\usepackage{indentfirst}
\usepackage{feynmf}   
\usepackage{slashed}  
\usepackage{cases}
\usepackage{color}
\usepackage{multirow}
\usepackage{graphicx,color,dcolumn,booktabs,bm}

\graphicspath{{Figures/}} %

\graphicspath{{Figures/}} %

\usepackage[colorlinks, citecolor=blue,anchorcolor=red,menucolor=red, linkcolor=red,filecolor=red,runcolor=red,urlcolor=blue,frenchlinks=red]{hyperref}

\allowdisplaybreaks

\begin{document}

\title{Possible hidden-bottom molecular pentaquarks from $P$-wave $\Lambda_bB^{(*)}/\Sigma_b^{(*)}B^{(*)}$ interactions}

\author{Yu-Xin Wan$^{1}$}
\author{Rui Chen$^{1,2}$}\email{chenrui@hunnu.edu.cn}
\author{Fu-Lai Wang$^{3}$}
\author{Qi Huang$^{4}$}
\affiliation{
$^1$Key Laboratory of Low-Dimensional Quantum Structures and Quantum Control of Ministry of Education, Department of Physics and Synergetic Innovation Center for Quantum Effects and Applications, Hunan Normal University, Changsha 410081, China\\
$^2$Hunan Research Center of the Basic Discipline for Quantum Effects and Quantum Technologies, Hunan Normal University, Changsha 410081, China\\
$^3$School of Physical Science and Technology, Lanzhou University, Lanzhou 730000, China\\
$^4$School of Physical Science and Technology, Nanjing Normal University, Nanjing 210023, China}
\date{\today}

\begin{abstract}

In this work, we perform a systematic investigation of the hidden-bottom molecular pentaquark states, encompassing both bound states and resonances, which originate from the $P$-wave interactions between ground-state bottom baryons ($\Lambda_b$, $\Sigma_b^{(*)}$) and ground-state antibottom mesons ($B^{(*)}$). Adopting the one-boson-exchange model and including the coupled-channel effects, we derive the effective potentials for all allowed quantum numbers $I(J^P) = 1/2(1/2^+)$, $1/2(3/2^+)$, $1/2(5/2^+)$, $1/2(7/2^+)$, $3/2(1/2^+)$, $3/2(3/2^+)$, $3/2(5/2^+)$, and $3/2(7/2^+)$. We then solve the coupled-channel Schr\"odinger equations to search for the bound-state solutions and perform the phase-shift analyses to identify resonance poles. Our results reveal a rich spectrum of positive-parity hidden-bottom molecular pentaquark candidates. In the isospin $I=1/2$ sector, we find several loosely bound states and associated resonances, particularly in the $\Sigma_b B^*$ and $\Sigma_b^* B^*$ channels, where the coupled-channel dynamics plays an essential role in their formation. In the isospin $I=3/2$ sector, the attraction is generally weaker because of the isospin factors. Nevertheless, we still obtain the loosely bound states and resonances, such as the $\Sigma_b^* B^*$ states with $J^P=3/2^+$, $5/2^+$, and $7/2^+$. The prominence of high-spin partial waves, for instance the $^6P_J$ components, underscores the importance of the spin-spin and tensor interactions. Our predictions provide a comprehensive and systematic spectrum of the $P$-wave hidden-bottom molecular pentaquark states and offer clear guidance for future experimental searches at LHCb and Belle~II.

\end{abstract}


\maketitle

\section{Introduction}

The conventional quark model has long served as a successful framework for describing most hadrons as the mesons ($q\bar{q}$) and baryons ($qqq$). However, as the fundamental theory of strong interactions, Quantum Chromodynamics (QCD) does not inherently exclude the existence of more complex quark configurations. Over the past two decades, with the increasing accumulation of high-energy experimental data from facilities such as Belle, BESIII, and LHCb, a growing number of candidate exotic hadronic states beyond the conventional quark model have been reported \cite{Liu:2013waa,Hosaka:2016pey,Chen:2016qju,Richard:2016eis,Lebed:2016hpi,Olsen:2017bmm,Guo:2017jvc,Liu:2019zoy,Brambilla:2019esw,Meng:2022ozq,Chen:2022asf,Wang:2025dur,Bai:2026atm}. These discoveries have propelled the study of exotic hadrons to the forefront of hadron physics.

Among these exotic states, the hadronic molecular picture has attracted considerable attention. In this picture, two or more hadrons are proposed to form weakly bound nucleus-like structures via the residual strong interaction. This interpretation is often invoked to explain near-threshold hadronic states such as the $XYZ$, $P_c$, $P_{cs}$, and $T_{cc}$ families. Most current studies have concentrated on meson-meson and meson-baryon systems in the charm sector. For instance, the $X(3872)$ \cite{Belle:2003nnu} and the $T_{cc}^+(3875)$ \cite{LHCb:2021vvq,LHCb:2021auc} are frequently interpreted as $D\bar{D}^*$ \cite{Liu:2008fh,Thomas:2008ja,Lee:2009hy,Li:2012cs,Sun:2012zzd,Zhao:2014gqa,Lin:2024qcq,Lu:2025zae,BaBar:2006qlj,Shen:2024npc,Esposito:2021vhu,Wang:2024ytk,Liu:2009qhy} and $DD^*$ \cite{Chen:2021vhg,Albaladejo:2021vln,Du:2021zzh,Moradpouri:2025psk,Du:2025vkm,Collins:2024sfi,Abolnikov:2024key,Li:2012ss,Xu:2017tsr,Ling:2021bir,Dong:2021bvy,Feijoo:2021ppq,Ren:2021dsi,Chen:2021cfl,Santowsky:2021bhy,Deng:2021gnb,Kamiya:2022thy,Abreu:2022sra,Chen:2022vpo} molecules, respectively. Similarly, the $P_c(4312)$, $P_c(4440)$, and $P_c(4457)$ states \cite{Aaij:2019vzc} have often been assigned as $\Sigma_c\bar{D}^{(*)}$ molecular pentaquarks \cite{Wu:2010jy,Yang:2011wz,Wang:2011rga,Wu:2012md,Shen:2019evi,Guo:2019fdo,Xiao:2019mvs,He:2019ify,Uchino:2015uha,Chen:2019bip,Yamaguchi:2019seo,Burns:2019iih,Meng:2019ilv,PavonValderrama:2019nbk,Du:2019pij,Wang:2019ato,Xu:2025mhc,Chen:2019asm}.

The extensive investigation of the hadronic molecules in the charm sector naturally prompts the question of whether analogous states also exist in the bottom sector. Compared with their charmed counterparts, studies of bottom hadronic molecules have progressed more slowly, partly due to the higher production thresholds and the scarcity of experimental data in earlier years. Nevertheless, the ongoing advances in experimental capabilities are expected to create new opportunities for discovering bottom molecular states. In this sense, research on bottom hadronic molecules is likely entering a phase of rapid growth. Indeed, as early as 2011, two charged hidden-bottom structures, $Z_b(10610)$ and $Z_b(10650)$, were observed by the Belle collaboration \cite{Belle:2011aa}. Although various theoretical interpretations have been proposed for these states, their proximity to the $B\bar{B}^*$ and $B^*\bar{B}^*$ thresholds renders the hidden-bottom molecular tetraquark interpretation particularly attractive \cite{Prelovsek:2019ywc, Zhang:2011jja, Yang:2011rp, Nieves:2011zz, Sun:2011uh, Cleven:2011gp, Mehen:2011yh, Ohkoda:2011vj, Li:2012wf, Liu:2017mrh, Zhao:2014gqa, Chen:2015ata, Wang:2018jlv, Bondar:2011ev, Dong:2012hc, Li:2012uc, Wu:2020edh, Xiao:2017uve, Wu:2022hck, Wu:2018xaa}.

From the theoretical perspective, the one-boson-exchange (OBE) model has become a powerful tool for describing the hadron-hadron interactions at low energies. It successfully accounts for fundamental phenomena such as nucleon binding and nucleon-nucleon scattering at the hadronic level, and has been extensively employed in studies of charmed hadronic interactions. In the hidden-bottom sector, several investigations have predicted the existence of possible bound states or resonances in the $S$-wave $\Lambda_bB^{(*)}$ and $\Sigma_b^{(*)}B^{(*)}$ systems, drawing on the guidance of the heavy-quark spin symmetry and the experience accumulated in the charm sector \cite{Chen:2015loa,Zhu:2020vto,He:2019rva,Karliner:2015ina}.

Compared with the well-studied $S$-wave interactions, the $P$-wave interactions in bottom baryon-antibottom meson systems have received considerably less attention. The main reason for this situation is that relative orbital angular momentum $L=1$ will give rise to a centrifugal barrier, which is usually considered to suppress the formation of the loosely bound states. However, a series of works indicate that such an interaction actually can generate resonances near the thresholds \cite{Wang:2025kpm,Cui:2025elw,Yamaguchi:2011qw,Ohkoda:2012hv,Ohkoda:2011vj,Wang:2023ivd,Wang:2025jec,Wang:2024ukc,Chen:2025gxe}. Moreover, for a $P$-wave molecule, the orbital excitation flips the overall parity, leading to quantum numbers distinct from those of the $S$-wave systems. In view of these considerations, a comprehensive investigation of the $P$-wave interactions in bottom baryon-antibottom meson systems is both timely and well motivated.

In this work, we investigate the $P$-wave interactions between ground-state bottom baryons ($\Lambda_b$, $\Sigma_b^{(*)}$) and ground-state antibottom mesons ($B^{(*)}$). We adopt the OBE model and incorporate the coupled-channel effects to derive the effective potentials. With these potentials, we solve the coupled-channel Schr\"odinger equations to search for the bound-state solutions and perform the phase-shift analyses for the systems under consideration. Our primary goal is to identify possible hidden-bottom bound pentaquark states or resonances with positive parity in the $P$-wave $\Lambda_bB^{(*)}/\Sigma_b^{(*)}B^{(*)}$ configurations. Through this study, we establish, for the first time, a systematic spectrum of the $P$-wave hidden-bottom molecular pentaquarks from the ground state to the orbital excited states. Our results provide clear guidance for experimental searches at facilities such as LHCb and Belle II, and offer key theoretical insights into the interaction mechanisms of the heavy-flavor baryon-meson systems. Moreover, the experimental identification of these possible hidden-bottom bound molecular pentaquarks may also help to clarify the internal structures of the observed $P_c$ states \cite{Aaij:2019vzc}.

The remainder of this paper is organized as follows. In Sec.~\ref{sec2}, we present the OBE effective potentials for the $P$-wave $\Lambda_b{B}^{(*)}/\Sigma_b^{(*)}{B}^{(*)}$ systems. The numerical results are then discussed in Sec.~\ref{sec3}. Finally, a brief summary is given in Sec.~\ref{sec4}.

\section{OBE effective potentials for the $P$-wave $\Lambda_b{B}^{(*)}/\Sigma_b^{(*)}{B}^{(*)}$ systems}\label{sec2}

Prior to deriving the OBE effective potentials for the $P$-wave $\Lambda_bB^{(*)}/\Sigma_b^{(*)}B^{(*)}$ systems, we first construct their isospin and spin-orbit wave functions. The isospin wave functions $|I, I_3\rangle$ for the aforementioned systems are given as follows:
\begin{eqnarray}&&\left\{\begin{array}{cc}
\left|\frac{1}{2},\frac{1}{2}\right\rangle =
     \left|\Lambda_b^{0}{B}^{(*)+}\right\rangle\\
\left|\frac{1}{2},-\frac{1}{2}\right\rangle =\left|\Lambda_b^{0}{B}^{(*)0}\right\rangle
     \end{array}\right.,\\
     &&\left\{\begin{array}{c}
\left|\frac{1}{2},\frac{1}{2}\right\rangle =
     \sqrt{\frac{2}{3}}\left|\Sigma_b^{(*)+}{B}^{(*)0}\right\rangle
     -\frac{1}{\sqrt{3}}\left|\Sigma_b^{(*)0}{B}^{(*)+}\right\rangle\\
\left|\frac{1}{2},-\frac{1}{2}\right\rangle =
     \frac{1}{\sqrt{3}}\left|\Sigma_b^{(*)0}{B}^{(*)0}\right\rangle
     -\sqrt{\frac{2}{3}}\left|\Sigma_b^{(*)-}{B}^{(*)+}\right\rangle
     \end{array}\right.,\\
&&\left\{\begin{array}{l}
\left|\frac{3}{2},\frac{3}{2}\right\rangle = \left|\Sigma_b^{(*)+}{B}^{(*)+}\right\rangle\\
\left|\frac{3}{2},\frac{1}{2}\right\rangle =
     \frac{1}{\sqrt{3}}\left|\Sigma_b^{(*)+}{B}^{(*)0}\right\rangle+\sqrt{\frac{2}{3}}
     \left|\Sigma_b^{(*)0}{B}^{(*)+}\right\rangle\\
\left|\frac{3}{2},-\frac{1}{2}\right\rangle =\sqrt{\frac{2}{3}}
    \left|\Sigma_b^{(*)0}{B}^{(*)0}\right\rangle+ \frac{1}{\sqrt{3}}\left|\Sigma_b^{(*)-}{B}^{(*)+}\right\rangle\\
\left|\frac{3}{2},-\frac{3}{2}\right\rangle =
     \left|\Sigma_b^{(*)-}{B}^{(*)0}\right\rangle
     \end{array}\right..
\end{eqnarray}
Here, $I$ and $I_3$ denote the isospin and its third component for the system under consideration, respectively.

For the $P$-wave $\Lambda_bB^{(*)}/\Sigma_b^{(*)}B^{(*)}$ systems, the allowed spin-parity quantum numbers are $J^P=1/2^+$, $3/2^+$, $5/2^+$, and $7/2^+$. In Table \ref{channel}, we summarize the possible channels involved in our calculations, where the notation $|{}^{2S+1}L_J\rangle$ denotes the spin-orbit wave functions. The general expressions of the spin-orbit wave functions for the $\Lambda_b{B}^{(*)}/\Sigma_b^{(*)}{B}^{(*)}$ systems are constructed as
\begin{eqnarray}
\Lambda_b(\Sigma_b)B:\,\,\left|{}^{2S+1}L_{J}\right\rangle &=&
\sum_{m_S,m_L}C^{J,M}_{\frac{1}{2}m_S,Lm_L}
          \chi_{\frac{1}{2}m_S}|Y_{L,m_L}\rangle,\nonumber\\
\Sigma_b^*{B}:\,\, \left|{}^{2S+1}L_{J}\right\rangle &=&
\sum_{m_S,m_L}C^{J,M}_{\frac{3}{2}m_S,Lm_L}
          \Phi_{\frac{3}{2}m_S}|Y_{L,m_L}\rangle,\nonumber\\
\Lambda_b(\Sigma_b){B}^*: \left|{}^{2S+1}L_{J}\right\rangle &=&
\sum_{m,m'}^{m_S,m_L}C^{S,m_S}_{\frac{1}{2}m,1m'}C^{J,M}_{Sm_S,Lm_L}
          \chi_{\frac{1}{2}m}\epsilon^{m'}|Y_{L,m_L}\rangle,\nonumber\\
\Sigma_b^*{B}^*: \left|{}^{2S+1}L_{J}\right\rangle &=&
\sum_{m,m'}^{m_S,m_L}C^{S,m_S}_{\frac{3}{2}m,1m'}C^{J,M}_{Sm_S,Lm_L}
          \Phi_{\frac{3}{2}m}\epsilon^{m'}|Y_{L,m_L}\rangle.\nonumber
\end{eqnarray}
Here, $C^{J,M}_{\frac{1}{2}m_S,Lm_L}$, $C^{J,M}_{Sm_S,Lm_L}$, $C^{S,m_S}_{\frac{1}{2}m,1m'}$, and $C^{S,m_S}_{\frac{3}{2}m,1m'}$ are the Clebsch-Gordan coefficients. $\chi_{\frac{1}{2}m}$ and $Y_{L,m_L}$ stand for the spin wave function and the spherical harmonic functions, respectively. The polarization vector $\epsilon$ for the ${B}^*$ vector meson is defined as $\epsilon_{\pm}^{m}=\mp\frac{1}{\sqrt{2}}\left(\epsilon_x^{m}{\pm}i\epsilon_y^{m}\right)$ and $\epsilon_0^{m}=\epsilon_z^{m}$, while these components satisfy $\epsilon_{\pm1}= \frac{1}{\sqrt{2}}\left(0,\pm1,i,0\right)$ and $\epsilon_{0} =\left(0,0,0,-1\right)$. The polarization tensor $\Phi_{\frac{3}{2}m}$ for the $\Sigma_b^*$ baryon is expressed as $\Phi_{\frac{3}{2}m}=\sum_{m_1,m_2}\langle\frac{1}{2},m_1;1,m_2|\frac{3}{2},m\rangle\chi_{\frac{1}{2} m_1}\epsilon^{m_2}$.

\renewcommand\tabcolsep{0.12cm}
\renewcommand{\arraystretch}{1.8}
\begin{table}[!htbp]
\centering
\caption{Possible channels involved in our calculations.}\label{channel}
\begin{tabular}{l|l}
\toprule[1pt]\toprule[1pt]
$I(J^P)$ & {Channels} \\
\hline
\multirow{2}{*}{$\frac{1}{2}(\frac{1}{2}^+)$} &$\Lambda_b B$: $| ^2P_\frac{1}{2}\rangle$, ~$\Lambda_b B^*$: $| ^2P_\frac{1}{2}/{}^4P_\frac{1}{2}\rangle$,~  $\Sigma_b B$: $| ^2P_\frac{1}{2}\rangle$,~  $\Sigma_b^* B$: $| ^4P_\frac{1}{2}\rangle$, \\
&  $\Sigma_b B^*$: $| ^2P_\frac{1}{2}/ ^4P_\frac{1}{2}\rangle$,~ $\Sigma_b^* B^*$: $| ^2P_\frac{1}{2}/ {}^4P_\frac{1}{2}\rangle$. \\
\hline
\multirow{2}{*}{$\frac{3}{2}(\frac{1}{2}^+)$} & $\Sigma_b B$: $| ^2P_\frac{1}{2}\rangle$,~  $\Sigma_b^* B$: $| ^4P_\frac{1}{2}\rangle$, \\
&$\Sigma_b B^*$: $| ^2P_\frac{1}{2}/ ^4P_\frac{1}{2}\rangle$,~ $\Sigma_b^* B^*$: $| ^2P_\frac{1}{2}/ ^4P_\frac{1}{2}\rangle$. \\
\hline
\multirow{2}{*}{$\frac{1}{2}(\frac{3}{2}^+)$} &$\Lambda_b B$: $| ^2P_\frac{3}{2}\rangle$,~  $\Lambda_b B^*$: $| ^2P_\frac{3}{2}/ ^4P_\frac{3}{2}\rangle$,~
$\Sigma_b B$: $| ^2P_\frac{3}{2}\rangle$,~  $\Sigma_b^* B$: $| ^4P_\frac{3}{2}\rangle$,\\   
&$\Sigma_b B^*$: $| ^2P_\frac{3}{2}/ ^4P_\frac{3}{2}\rangle$,~ $\Sigma_b^* B^*$: $| ^2P_\frac{3}{2}/ ^4P_\frac{3}{2}/ ^6P_\frac{3}{2}\rangle$. \\
\hline
\multirow{2}{*}{$\frac{3}{2}(\frac{3}{2}^+)$} & $\Sigma_b B$: $| ^2P_\frac{3}{2}\rangle$,~  $\Sigma_b^* B$: $| ^4P_\frac{3}{2}\rangle$,~ $\Sigma_b B^*$: $| ^2P_\frac{3}{2}/ ^4P_\frac{3}{2}\rangle$,\\ 
& $\Sigma_b^* B^*$: $| ^2P_\frac{3}{2}/ ^4P_\frac{3}{2}/ ^6P_\frac{3}{2}\rangle$. \\
\hline
$\frac{1}{2}(\frac{5}{2}^+)$  &$\Lambda_b B^*$: $| ^4P_\frac{5}{2}\rangle$,~ $\Sigma_b^* B$: $| ^4P_\frac{5}{2}\rangle$,~
$\Sigma_b B^*$: $| ^4P_\frac{5}{2}\rangle$,~ $\Sigma_b^* B^*$: $| ^4P_\frac{5}{2}/ ^6P_\frac{5}{2}\rangle$.\\
\hline
$\frac{3}{2}(\frac{5}{2}^+)$ & $\Sigma_b^* B$: $| ^4P_\frac{5}{2}\rangle$,~  $\Sigma_b B^*$: $| ^4P_\frac{5}{2}\rangle$,~$\Sigma_b^* B^*$: $| ^4P_\frac{5}{2}/ ^6P_\frac{5}{2}\rangle$.\\
\hline
$\frac{1}{2}(\frac{7}{2}^+)$  &$\Sigma_b^* B^*$: $| ^6P_\frac{7}{2}\rangle$.\\
\hline
$\frac{3}{2}(\frac{7}{2}^+)$  &$\Sigma_b^* B^*$: $| ^6P_\frac{7}{2}\rangle$.\\
\bottomrule[1pt]\bottomrule[1pt]
\end{tabular}
\end{table}

In the OBE model, two hadrons interact via the exchange of a virtual particle, namely a boson, which mediates the interaction force. For the $\Lambda_b{B}^{(*)}/\Sigma_b^{(*)}{B}^{(*)}$ systems, the exchanged mesons include the scalar meson $\sigma$, the pseudoscalar mesons $\pi,\eta$, and vector mesons $\rho,\omega$. According to the heavy-quark symmetry and chiral symmetry \cite{Yan:1992gz,Wise:1992hn,Casalbuoni:1996pg,Falk:1992cx,Liu:2011xc}, the effective Lagrangians describing the interactions between the heavy mesons/baryons and the light mesons can be constructed as follows:
\begin{align}
\mathcal{L}_{H}
&= g_S\langle \bar{H}^{(\overline{Q})}_a\sigma H^{(\overline{Q})}_b\rangle
   + ig\langle \bar{H}^{(\overline{Q})}_a\gamma_{\mu}{\mathcal A}_{ab}^{\mu}\gamma_5H^{(\overline{Q})}_b\rangle \notag \\
&\quad - i\beta\langle \bar{H}^{(\overline{Q})}_a v_{\mu}\left(\mathcal{V}^{\mu}-\rho^{\mu}\right)_{ab}H^{(\overline{Q})}_b\rangle \notag \\
&\quad + i\lambda\langle \bar{H}^{(\overline{Q})}_a\sigma_{\mu\nu}F^{\mu\nu}(\rho)H^{(\overline{Q})}_b\rangle, \label{lag1} \\
\mathcal{L}_{\mathcal{B}_{\bar{3}}}
&= l_B\langle\bar{\mathcal{B}}_{\bar{3}}\sigma\mathcal{B}_{\bar{3}}\rangle
   + i\beta_B\langle\bar{\mathcal{B}}_{\bar{3}}v^{\mu}(\mathcal{V}_{\mu}-\rho_{\mu})\mathcal{B}_{\bar{3}}\rangle, \\
\mathcal{L}_{\mathcal{B}^{(*)}_6}
&= l_S\langle\bar{\mathcal{S}}_{\mu}\sigma\mathcal{S}^{\mu}\rangle
   - \frac{3}{2}g_1\varepsilon^{\mu\nu\lambda\kappa}v_{\kappa}\langle\bar{\mathcal{S}}_{\mu}{\mathcal A}_{\nu}\mathcal{S}_{\lambda}\rangle \notag \\
&\quad + i\beta_{S}\langle\bar{\mathcal{S}}_{\mu}v_{\alpha}\left(\mathcal{V}^{\alpha}-\rho^{\alpha}\right) \mathcal{S}^{\mu}\rangle \notag \\
&\quad + \lambda_S\langle\bar{\mathcal{S}}_{\mu}F^{\mu\nu}(\rho)\mathcal{S}_{\nu}\rangle, \\
\mathcal{L}_{\mathcal{B}_{\bar{3}}\mathcal{B}^{(*)}_6}
&= ig_4\langle\bar{\mathcal{S}^{\mu}}{\mathcal A}_{\mu}\mathcal{B}_{\bar{3}}\rangle
   + i\lambda_I\varepsilon^{\mu\nu\lambda\kappa}v_{\mu}\langle \bar{\mathcal{S}}_{\nu}F_{\lambda\kappa}\mathcal{B}_{\bar{3}}\rangle + \text{h.c.}. \label{lag2}
\end{align}
Here, the multiplet fields $H^{(\overline{Q})}$ and $\mathcal{S}$ are the linear combinations of the $S$-wave charmed mesons and the $S$-wave charmed baryons in the $6_F$ flavor representation, respectively. $H^{(\overline{Q})} = \left[\tilde{\mathcal{P}}^{*\mu}\gamma_{\mu}-\tilde{\mathcal{P}}\gamma_5\right]\frac{1-\rlap\slash v}{2}$ with $\tilde{\mathcal{P}}=\left({B}^+,\,B^0\right)^T$ and $\tilde{\mathcal{P}}^*=\left({B}^{*+},\,B^{*0}\right)^T$. $\mathcal{S}_{\mu} =
-\sqrt{\frac{1}{3}}(\gamma_{\mu}+v_{\mu})\gamma^5\mathcal{B}_6+\mathcal{B}_{6\mu}^*$. $\mathcal{A}_{\mu}$ and $\mathcal{V}_{\mu}$ correspond to the axial current and vector current, respectively, $\mathcal{A}_{\mu} = \frac{1}{2}(\xi^{\dag}\partial_{\mu}\xi-\xi\partial_{\mu}\xi^{\dag})=\frac{i}{f_{\pi}}
\partial_{\mu}\mathbb{P}+\ldots$ and $\mathcal{V}_{\mu} =
\frac{1}{2}(\xi^{\dag}\partial_{\mu}\xi-\xi\partial_{\mu}\xi^{\dag})
=\frac{i}{2f_{\pi}^2}\left[\mathbb{P},\partial_{\mu}\mathbb{P}\right]+\ldots$ with $\xi=\text{exp}(i\mathbb{P}/f_{\pi})$. $\rho_{ba}^{\mu}=ig_V\mathbb{V}_{ba}^{\mu}/\sqrt{2}$, $F^{\mu\nu}(\rho)=\partial^{\mu}\rho^{\nu}-\partial^{\nu}\rho^{\mu}
+\left[\rho^{\mu},\rho^{\nu}\right]$. $\mathbb{P}$ and $\mathbb{V}$ stand for the pseudoscalar and vector meson matrixes, respectively. The matrices for $\mathcal{B}_{\bar{3}}$, $\mathcal{B}_6^{(*)}$, $\mathbb{P}$, and $\mathbb{V}$ are expressed as
\begin{eqnarray*}\left.\begin{array}{ll}
\mathcal{B}_{\bar{3}} = {\left(\begin{array}{ccc}
         0    &\Lambda_b^0       \\
        -\Lambda_b^0      &0    
                \end{array}\right),}
&\mathcal{B}_6^{(*)} = \left(\begin{array}{cc}
         \Sigma_b^{{(*)}+}              &\frac{1}{\sqrt{2}}\Sigma_b^{{(*)}0}\\
         \frac{1}{\sqrt{2}}\Sigma_b^{{(*)}0}      &\Sigma_b^{{(*)}-}
\end{array}\right),\\
\mathbb{P} = \left(\begin{array}{cc}
\frac{\pi^0}{\sqrt{2}}+\frac{\eta}{\sqrt{6}} &\pi^+\\
\pi^- &-\frac{\pi^0}{\sqrt{2}}+\frac{\eta}{\sqrt{6}}
\end{array}\right),
&\mathbb{V} = \left(\begin{array}{cc}
\frac{\rho^0}{\sqrt{2}}+\frac{\omega}{\sqrt{2}}  &\rho^+\\
\rho^- &-\frac{\rho^0}{\sqrt{2}}+\frac{\omega}{\sqrt{2}}
\end{array}\right),\end{array}\right.
\end{eqnarray*}
respectively. Additionally, the normalization relations for the scalar meson, the vector meson, and the heavy baryons are
\begin{eqnarray*}
\langle0|\tilde{\mathcal{P}}|\bar{Q}q(0^-)\rangle &=& \sqrt{M_{\tilde{\mathcal{P}}}}, \\
\langle0|\tilde{\mathcal{P}}^*_{\mu}|\bar{Q}q(1^-)\rangle &=& \epsilon_{\mu}\sqrt{M_{\tilde{\mathcal{P}}^*}}, \\
\langle0|\mathcal{B}_6|Qqq(1/2^+)\rangle &=& \sqrt{2M_{\mathcal{B}_6}}\left(\left(1-\frac{\bm{p}^2}{8M_{\mathcal{B}_6}^2}\right)
\chi,\frac{\bm{\sigma}\cdot\bm{p}}{2M_{\mathcal{B}_6}}\chi\right)^T,\\
\langle0|\mathcal{B}_{6}^{*\mu}|Qqq(3/2^+)\rangle &=& \sum_{m_1,m_2}C_{1/2,m_1;1,m_2}^{3/2,m_1+m_2}
\sqrt{2M_{\mathcal{B}_{6}^*}}
\epsilon^{\mu}_{m_2}\nonumber\\
&&\times\left(\left(1-\frac{\bm{p}^2}{8M_{\mathcal{B}_6^*}}\right)
\chi_{\frac{1}{2},m_1},\frac{\bm{\sigma}\cdot\bm{p}}
{2M_{\mathcal{B}_6^*}}\chi_{\frac{1}{2},m_1}\right)^T.
\end{eqnarray*}
In this work, we adopt the same values of the coupling constants as in Refs.~\cite{Liu:2011xc,Casalbuoni:1996pg,Falk:1992cx,Yang:2011wz}, which are summarized in Table~\ref{coupling}. The relative phases between these coupling constants are fixed by the quark model \cite{Liu:2011xc}.

\renewcommand\tabcolsep{0.4cm}
\renewcommand{\arraystretch}{1.8}
\begin{table}[!htbp]
  \centering
  \caption{Coupling constants are taken from Refs.
\cite{Liu:2011xc,Casalbuoni:1996pg,Falk:1992cx,Yang:2011wz}.} \label{coupling}
    \begin{tabular}{lll}
    \toprule[1pt]\toprule[1pt]
    $g_S=0.76$       &$g=0.59$     &$\beta=0.9$\\
    $l_S=6.2$        &$g_1=0.94$   &$\beta_S=-1.74$\\
    $\lambda=0.56$ GeV$^{-1}$     &$\lambda_S=-3.31$ GeV$^{-1}$   &$g_V=5.8$\\ 
  $f_{\pi}=132 $ MeV    &$g_4=1.06$    &$l_B=-3.65$\\$\lambda_Ig_V=6.8$ GeV$^{-1}$     &$\beta_B g_V=6$\\
  \bottomrule[1pt]\bottomrule[1pt]
  \end{tabular}
\end{table}

Using the above effective Lagrangians, one can write down the OBE scattering amplitude $\mathcal{M}[B_1\bar{M}_2\to B_3\bar{M}_4]$ for the $B_1\bar{M}_2\to B_3\bar{M}_4$ process. In the Breit approximation \cite{Berestetskii:1982qgu}, the effective potential in momentum space can be related to the corresponding scattering amplitude as $\mathcal{V}(\bm{q})=-\mathcal{M}[B_1\bar{M}_2\to B_3\bar{M}_4]/\sqrt{16m_{B_1}m_{\bar{M}_2}m_{B_3}m_{\bar{M}_4}}$. To obtain the OBE effective potential in coordinate space, we further perform a Fourier transformation, namely
\begin{eqnarray*}
\mathcal{V}(\bm{r}) &=&
          \int\frac{d^3\bm{q} }{(2\pi)^3}e^{i\bm{q}\cdot\bm{r}}
          \mathcal{V}(\bm{q})\mathcal{F}^2(q^2,m_E^2).
\end{eqnarray*}
Here, we introduce a monopole form factor at every interaction vertex to compensate for the off-shell effects of the exchanged meson, which takes the form $\mathcal{F}^2(q^2,m_E^2)=(\Lambda^2-m_E^2)/(q^2-m_E^2)$, where $\Lambda$, $m_E$, and $q$ stand for the cutoff, mass, and four-momentum of the exchanged particle, respectively. As a phenomenological parameter, the cutoff $\Lambda$ is often varied in the range of $1$ to $2$ GeV in the quantitative description of deuteron properties and nucleon-nucleon scattering data (see the review in Ref.~\cite{Machleidt:1987hj}). In the following, we adopt this criterion to analyze our results.

\renewcommand\tabcolsep{0.28cm}
\renewcommand{\arraystretch}{1.8}
\begin{table*}[!htbp]
  \centering
  \caption{The detailed expressions of the $I=1/2$ or $3/2$ OBE effective potentials for all discussed scattering processes. Here, $\mathcal{G}$ is the isospin factor, taken as $-1$ for the isospin-$1/2$ system and $1/2$ for the isospin-$3/2$ system. And we define several functions: $Y_{\Lambda,m} =\frac{1}{4\pi r}(e^{-mr}-e^{-\Lambda r})-\frac{\Lambda^2-m^2}{8\pi \Lambda}e^{-\Lambda r}$, $\mathcal{Y}^{ij}_{\Lambda, m_a}=\mathcal{D}_{ij}Y_{\Lambda,m}$, $\mathcal{Z}^{ij}_{\Lambda, m_a}=\left(\mathcal{E}_{ij}\nabla^2+\mathcal{F}_{ij}r\frac{\partial}{\partial r}\frac{1}{r}\frac{\partial}{\partial r}\right)Y_{\Lambda,m}$, and $\mathcal{Z}^{\prime ij}_{\Lambda, m_a}=\left(2\mathcal{E}_{ij}\nabla^2-\mathcal{F}_{ij}r\frac{\partial}{\partial r}\frac{1}{r}\frac{\partial}{\partial r}\right)Y_{\Lambda,m}$. $\mathcal{D}_{ij}$, $\mathcal{E}_{ij}$, and $\mathcal{F}_{ij}$ denote the spin-spin interaction and the tensor operators, respectively. Their concrete expressions and the corresponding matrix elements are presented in Appendix~\ref{app}. The quantities $\Lambda_i$ and $m_i$ appearing in the potentials are defined by $\Lambda_i^2 = \Lambda^2 - q_i^2$ and $m_i^2 = m^2 - q_i^2$, with $i=1,\ldots,12$. The momentum shift $q_i^2$ is given by $q_i^2 = \left(\frac{M_1^2+M_4^2-M_2^2 -M_3^2}{2(M_1+M_2)}\right)^2$, where $M_1$ and $M_2$ are the masses of the initial baryon and meson, respectively, and $M_3$ and $M_4$ are those of the final baryon and meson.}\label{potentials}
  \begin{tabular}{c|l||c|l}
  \toprule[1pt]\toprule[1pt]
  Processes    &OBE effective potentials    &Processes    &OBE effective potentials\\\hline
   $\Lambda_b B\to\Lambda_b B$
      &$2g_s l_B Y_{\Lambda,m_{\sigma}} + \frac{1}{2} \beta \beta_B g_V^2 Y_{\Lambda,m_{\omega}}$
 & $\Lambda_b B\to\Sigma_b B^*$
      &$-\frac{gg_4}{3\sqrt{2}f_\pi^2}\mathcal{Z}^{13}_{\Lambda_1,m_{\pi1}}+\frac{2\lambda\lambda_Ig_V^2}{3\sqrt{2}}\mathcal{Z}^{13}_{\Lambda_1,m_{\rho1}}$\\\hline
  $\Lambda_b B\to\Sigma_b^* B^*$
      &$\frac{gg_4}{\sqrt{6}f_\pi^2}\mathcal{Z}^{14}_{\Lambda_2,m_{\pi2}}+\frac{\lambda\lambda_Ig_V^2}{\sqrt{6}}\mathcal{Z}^{\prime14}_{\Lambda_2,m_{\rho2}}$
  &$\Lambda_b B^*\to\Lambda_b B^*$
      &$2g_s l_B \mathcal{Y}^{22}_{\Lambda,m_{\sigma}} + \frac{1}{2} \beta \beta_B g_V^2\mathcal{Y}^{22}_{\Lambda,m_{\omega}}$\\\hline
  $\Lambda_b B^*\to\Sigma_b B$
      &$\frac{g g_4}{3\sqrt{2}f_\pi^2}\mathcal{Z}^{23}_{\Lambda_2,m_{\pi2}}+\frac{2\lambda\lambda_Ig_V^2}{3\sqrt{2}}\mathcal{Z}^{\prime23}_{\Lambda_2,m_{\rho2}}$
&  $\Lambda_b B^*\to\Sigma_b^* B$
      &$\frac{g g_4}{\sqrt{6}f_\pi^2}\mathcal{Z}^{24}_{\Lambda_3,m_{\pi3}}-\frac{\lambda\lambda_Ig_V^2}{\sqrt{6}}\mathcal{Z}^{\prime24}_{\Lambda_3,m_{\rho3}}$\\\hline
  $\Lambda_b B^*\to\Sigma_b B^*$
      &$-\frac{g g_4}{3\sqrt{2}f_\pi^2}\mathcal{Z}^{25}_{\Lambda_4,m_{\pi4}}-\frac{2\lambda\lambda_Ig_V^2}{3\sqrt{2}}\mathcal{Z}^{\prime25}_{\Lambda_4,m_{\rho4}}$
&  $\Lambda_b B^*\to\Sigma_b^* B^*$
       &$\frac{g g_4}{\sqrt{6}f_\pi^2}\mathcal{Z}^{26}_{\Lambda_5,m_{\pi5}}+\frac{2\lambda\lambda_Ig_V^2}{\sqrt{6}}\mathcal{Z}^{\prime26}_{\Lambda_5,m_{\rho5}}$\\\hline
$\Sigma_b B\to\Sigma_b B$ &$-l_S g_S \mathcal{Y}^{33}_{\Lambda, m_{\sigma}} - \frac{\mathcal{G}}{2}\beta\beta_S g_V^2 \mathcal{Y}^{33}_{\Lambda, m_{\rho}} -\frac{1}{4}\beta\beta_S g_V^2 \mathcal{Y}^{33}_{\Lambda, m_{\omega}}$
&    $\Sigma_b B\to\Sigma_b^* B$
&$\frac{\mathcal{G}\beta\beta_Sg_V^2}{2\sqrt{3}}\mathcal{Y}^{34}_{\Lambda_3,m_{\rho3}}  +\frac{\beta\beta_Sg_V^2}{4\sqrt{3}}\mathcal{Y}^{34}_{\Lambda_3,m_{\omega3}}
    $\\\hline
  {$\Sigma_b B\to\Sigma_b B^*$}
     &
    $\frac{\mathcal{G}gg_1}{3f_\pi^2}\mathcal{Z}^{35}_{\Lambda_4,m_{\pi4}} +\frac{gg_1}{18f_\pi^2}\mathcal{Z}^{35}_{\Lambda_4,m_{\eta4}} +\frac{2\mathcal{G}\lambda\lambda_Sg_V^2}{9}\mathcal{Z}^{\prime35}_{\Lambda_4,m_{\rho4}} $
    &{$\Sigma_b B\to\Sigma_b^* B^*$}
     &
    $\frac{\mathcal{G}gg_1}{2\sqrt{3}f_\pi^2}\mathcal{Z}^{36}_{\Lambda_5,m_{\pi5}} +\frac{{gg_1}}{12\sqrt{3}{f_\pi^2}}\mathcal{Z}^{36}_{\Lambda_5,m_{\eta5}} +\frac{\mathcal{G}\lambda\lambda_Sg_V^2}{3\sqrt{3}}\mathcal{Z}^{\prime36}_{\Lambda_5,m_{\rho5}} $\\

    &$+\frac{\lambda\lambda_Sg_V^2}{9}\mathcal{Z}^{\prime35}_{\Lambda_4,m_{\omega4}}$ 
  
       &&$+\frac{\lambda\lambda_Sg_V^2}{6\sqrt{3}}\mathcal{Z}^{\prime36}_{\Lambda_5,m_{\omega5}}
$ \\\hline
  $\Sigma_b^* B\to\Sigma_b^* B$
 &
    $-l_Sg_S\mathcal{Y}^{44}_{\Lambda,m_{\sigma}} -\frac{\mathcal{G}\beta\beta_Sg_V^2}{2}\mathcal{Y}^{44}_{\Lambda,m_{\rho}} -\frac{\beta\beta_Sg_V^2}{4}\mathcal{Y}^{44}_{\Lambda,m_{\omega}}$

&{$\Sigma_b B^*\to\Sigma_b B^*$}
      &$-l_Sg_S\mathcal{Y}^{55}_{\Lambda,m_{\sigma}} +\frac{\mathcal{G}{gg_1}}{3{f_\pi^2}}\mathcal{Z}^{55}_{\Lambda,m_{\pi}} +\frac{{gg_1}}{18{f_\pi^2}}\mathcal{Z}^{55}_{\Lambda,m_{\eta}}$\\
      && &$ -\frac{\mathcal{G}\beta\beta_Sg_V^2}{2}\mathcal{Y}^{55}_{\Lambda,m_{\rho}} -\frac{2\mathcal{G}\lambda\lambda_Sg_V^2}{9}\mathcal{Z}^{\prime55}_{\Lambda,m_{\rho}} -\frac{\beta\beta_Sg_V^2}{4}\mathcal{Y}^{55}_{\Lambda,m_{\omega}}$ \\
      && &$-\frac{\lambda\lambda_Sg_V^2}{9}\mathcal{Z}^{\prime55}_{\Lambda,m_{\omega}}$\\\hline
   
  {$\Sigma_b^* B\to\Sigma_b B^*$}
      &
    $\frac{\mathcal{G}{gg_1}}{2\sqrt{3}{f_\pi^2}}\mathcal{Z}^{45}_{\Lambda_0,m_{\pi0}} +\frac{{gg_1}}{12\sqrt{3}{f_\pi^2}}\mathcal{Z}^{45}_{\Lambda_0,m_{\eta0}} -\frac{\mathcal{G}\lambda\lambda_Sg_V^2}{3\sqrt{3}}\mathcal{Z}^{\prime45}_{\Lambda_0,m_{\rho0}}$
&{$\Sigma_b^* B\to\Sigma_b^* B^*$}
      &$  \frac{\mathcal{G}{gg_1}}{2{f_\pi^2}}\mathcal{Z}^{46}_{\Lambda_1,m_{\pi1}} +\frac{{gg_1}}{12{f_\pi^2}}\mathcal{Z}^{46}_{\Lambda_1,m_{\eta1}} -\frac{\mathcal{G}\lambda\lambda_Sg_V^2}{3}\mathcal{Z}^{\prime46}_{\Lambda_1,m_{\rho1}}$\\
       &$ -\frac{\lambda\lambda_Sg_V^2}{6\sqrt{3}}\mathcal{Z}^{\prime45}_{\Lambda_0,m_{\omega0}}$ 
      & & $-\frac{\lambda\lambda_Sg_V^2}{6}\mathcal{Z}^{\prime 46}_{\Lambda_1,m_{\omega1}}$\\\hline
      
  {$\Sigma_b B^*\to\Sigma_b^* B^*$}
      &$\frac{l_Sg_S}{\sqrt{3}}\mathcal{Y}^{56}_{\Lambda_2,m_{\sigma2}} +\frac{\sqrt{3}\mathcal{G}{gg_1}}{6{f_\pi^2}}\mathcal{Z}^{56}_{\Lambda_2,m_{\pi2}} +\frac{\sqrt{3}{gg_1}}{36{f_\pi^2}}\mathcal{Z}^{56}_{\Lambda_2,m_{\eta2}}$ 
      & {$\Sigma_b^* B^*\to\Sigma_b^* B^*$}
        &$-l_Sg_S\mathcal{Y}^{66}_{\Lambda,m_{\sigma}} -\frac{\mathcal{G}{gg_1}}{2{f_\pi^2}}\mathcal{Z}^{66}_{\Lambda,m_{\pi}} -\frac{{gg_1}}{12{f_\pi^2}}\mathcal{Z}^{66}_{\Lambda,m_{\eta}}$\\
    
       &$+\frac{\mathcal{G}\beta\beta_Sg_V^2}{2\sqrt{3}}\mathcal{Y}^{56}_{\Lambda_2,m_{\rho2}} -\frac{\mathcal{G}\lambda\lambda_Sg_V^2}{3\sqrt{3}}\mathcal{Z}^{\prime56}_{\Lambda_2,m_{\rho2}} +\frac{\beta\beta_Sg_V^2}{4\sqrt{3}}\mathcal{Y}^{56}_{\Lambda_2,m_{\omega2}} $
       &&  $-\frac{\mathcal{G}\beta\beta_Sg_V^2}{2}\mathcal{Y}^{66}_{\Lambda,m_{\rho}} +\frac{\mathcal{G}\lambda\lambda_Sg_V^2}{3}\mathcal{Z}^{\prime66}_{\Lambda,m_{\rho}} -\frac{\beta\beta_Sg_V^2}{4}\mathcal{Y}^{66}_{\Lambda,m_{\omega}}$ \\
   
       &$-\frac{\lambda\lambda_Sg_V^2}{6\sqrt{3}}\mathcal{Z}^{\prime 56}_{\Lambda_2,m_{\omega2}}$
 &&$+\frac{\lambda\lambda_Sg_V^2}{6}\mathcal{Z}^{\prime66}_{\Lambda,m_{\omega}}$\\
  \bottomrule[1.0pt]\bottomrule[1.0pt]
  \end{tabular}
\end{table*}

In Table \ref{potentials}, we collect the OBE effective potentials for all discussed scattering processes. The relevant operators are summarized in Appendix \ref{app}.

\section{Numerical results}\label{sec3}

With the OBE effective potentials for the $P$-wave $\Lambda_b{B}^{(*)}/\Sigma_b^{(*)}{B}^{(*)}$ systems derived in the previous section, we next numerically solve the coupled-channel Schr\"odinger equation by varying the cutoff parameter $\Lambda$ of the effective potentials to search for possible hidden-bottom bound pentaquark states and resonances. We first solve the coupled-channel Schr\"odinger equation for the negative-energy case and look for shallow bound-state solutions within a reasonable cutoff range, namely $\Lambda \sim 1$ GeV. The corresponding binding energies are of the order of a few to tens of MeV, and the root-mean-square (RMS) radii of the systems are around 1 fm. The discussed systems that satisfy the characteristics of shallow bound states within the reasonable cutoff range can be considered as the promising molecular candidates.

In addition, we numerically solve the coupled-channel Schr\"odinger equation to compute the phase shifts for the systems under consideration. The physical boundary condition at the origin is taken as the regular boundary condition, while the asymptotic behavior of the reduced radial wave function defines the reaction matrix $K$, from which the scattering matrix $S$ and the phase shift $\delta_l(E)$ as functions of the center-of-mass energy $E$ can be extracted. It is well known that a resonance is identified when a phase shift crosses $\pi/2$ with a positive slope. This leads to the maximum of the scattering cross section, i.e., $\sigma_t=\frac{4\pi}{2\mu E}\sum_{l=0}^{\infty}(2l+1)\text{sin}^2\delta_l(E)$. The resonant width is defined as $\Gamma_r=2/\left(\frac{d\delta}{dE}\right)_{E_r}$. In the following, we analyze the results for each isospin and spin-parity sector for the $P$-wave $\Lambda_b{B}^{(*)}/\Sigma_b^{(*)}{B}^{(*)}$ systems.

We begin our investigation with the isospin-$1/2$ sector of the $\Lambda_bB^{(*)}/\Sigma_b^{(*)}B^{(*)}$ systems, where the OBE potentials are expected to produce sizable attraction due to the favorable isospin factor $\mathcal{G}=-1$. For each allowed total angular momentum and parity, we solve the coupled-channel Schr\"odinger equations to search for the bound-state solutions. We also perform a phase-shift analysis to identify possible resonance poles. In the following, we present our results for both the bound states and the resonances in this secto.

\subsection{$I(J^P)=1/2(1/2^+)$}

\paragraph{\bf{Bound states.}} We first study the $\Lambda_bB^{(*)}/\Sigma_b^{(*)}B^{(*)}$ coupled system with $I(J^P)=1/2(1/2^+)$ and search for possible hidden-bottom bound pentaquark candidates. In judging the reasonableness of the loosely bound state, we adhere to the criterion based on $\Lambda \sim 1$~GeV and require $E$ to be of the order of a few to several tens of MeV and $r_{\text{RMS}}$ to be around 1 fm. It is important to recognize that the binding energy $E$ extracted directly from the coupled-channel Schr\"odinger equations is defined relative to the lowest threshold among all included channels. In cases where the wave function is predominantly composed of a channel lying above this lowest threshold, the physically relevant binding energy with respect to the dominant channel, denoted as $\tilde E$, should be shifted accordingly $\tilde E = E + (M_{\text{low}} - M_{\text{dom}})$, where $M_{\text{low}}$ and $M_{\text{dom}}$ denote the masses of the lowest threshold and the dominant component, respectively. This shift can be quantitatively substantial, often yielding values of $\tilde E$ on the order of several hundred MeV. Such a large binding energy implies a compact spatial extent of the system, significantly smaller than $1\ \text{fm}$. Consequently, this type of bound state cannot be interpreted as a loosely bound hadronic molecule, in contrast to what a naive reading of the small $E$ might suggest \cite{Chen:2017xat}.

\renewcommand\tabcolsep{0.25cm}
\renewcommand{\arraystretch}{1.8}
\begin{table*}[!htbp]
\centering
\caption{The bound state solutions (the binding energy \(E\), the mass of the bound state $M$, the RMS radius \(r_{\text{RMS}}\), and probabilities \(p_i\)) for the $\Lambda_bB^{(*)}/\Sigma_b^{(*)}B^{(*)}$ coupled system with $I(J^P)=1/2(1/2^+)$. Here, $\Lambda$, $E$, $M$, and \(r_{\text{RMS}}\) are in unites of GeV, MeV, MeV, and fm, respectively.}\label{tab:couple1}
\begin{tabular}{ l l l l l l l l l l}
\toprule[1pt]\toprule[1pt]
 \(\Lambda\) & \(E\) &$M$ & \(r_{\text{RMS}}\) & \(\Lambda_b B(^2P_\frac{1}{2} )\) & \(\Lambda_b B^* (^2P_\frac{1}{2}/^4P_\frac{1}{2})\) & \(\Sigma_b B(^2P_\frac{1}{2})\)  & \(\Sigma_b^* B(^2P_\frac{1}{2})\) & \(\Sigma_b B^* (^2P_\frac{1}{2}/^4P_\frac{1}{2})\) & \(\Sigma_b^* B^* (^2P_\frac{1}{2}/^4P_\frac{1}{2})\) \\
\hline
  1.04&$-0.97$  &   10897.97
 &0.90 & \textbf{42.26} & 19.25/ 0.17&  5.42 & 1.01 & 3.36/ 4.96 &22.33/ 1.24\\
1.05 & $-10.05 $ &10888.89
&0.52 &\textbf{29.42}& 23.09/0.24 & 5.81 & 1.62 &  5.50 /4.97& 27.44/ 1.91 \\
 1.06 &$-20.99$  &10877.95
 &0.45 & {23.49}  &24.28 /0.46 & 5.72 & 2.03& 7.13/ 4.62&\textbf{29.92} / 2.35 
\\\bottomrule[1pt]\bottomrule[1pt]
 \end{tabular}
 \end{table*}

After solving the coupled-channel Schr\"odinger equations, we can obtain the bound-state solutions for the $\Lambda_bB^{(*)}/\Sigma_b^{(*)}B^{(*)}$ coupled system with $I(J^P)=1/2(1/2^+)$ as the cutoff $\Lambda$ is taken around 1 GeV. In Table \ref{tab:couple1}, we present the corresponding numerical results. Although we can obtain the bound-state solutions for this coupled system, the binding energy is somewhat sensitive to the cutoff $\Lambda$. In addition, the RMS radius deviates substantially from the typical size expected for a loosely bound molecular state. This deviation arises because the dominant channel is not the lowest mass threshold, namely $\Lambda_bB$, especially when the binding energy becomes relatively large \cite{Chen:2017xat}. As a consequence, this bound state cannot be regarded as a good molecular candidate. Nevertheless, the $\Lambda_bB^*$, $\Sigma_bB^*$, and $\Sigma_b^*B^*$ channels are found to play important roles in forming this bound state, as indicated by their large probability components. We therefore conclude that the interactions from these channels provide sufficiently strong attraction.

\renewcommand\tabcolsep{0.34cm}
\renewcommand{\arraystretch}{1.8}
\begin{table}[!htbp]
\centering
\caption{The bound state solutions (the binding energy \(E\), the mass of the bound state $M$, the RMS radius \(r_{\text{RMS}}\), and probabilities \(p_i\)) for the single $\Sigma_bB^*$ and $\Sigma_b^{*}B^*$ states with $I(J^P)=1/2(1/2^+)$. Here, $\Lambda$, $E$, $M$, and \(r_{\text{RMS}}\) are in unites of GeV, MeV, MeV, and fm, respectively.}\label{tab:single1}
\begin{tabular}{ l l l l l }
\toprule[1pt]\toprule[1pt]
 \(\Lambda\) & \(E\)  &$M$ & \(r_{\text{RMS}}\) &\(\Sigma_b B^* (^2P_\frac{1}{2}/^4P_\frac{1}{2})\)  \\
\cline{1-5}
 1.08 & $-2.18$   &11136.32
 & 1.22  &\textbf{88.13}/ 11.87
 \\
 1.10 & $-5.00$   &11133.50
& 0.98&\textbf{89.70}/ 10.30 
\\
  1.12& $-8.53$ &11129.97
 &0.84 &\textbf{90.89}/ 9.11 \\\cline{1-5}
{ \(\Lambda\)} & \(E\) &$M$ & \(r_{\text{RMS}}\) &\(\Sigma_b^* B^* (^2P_\frac{1}{2}/^4P_\frac{1}{2})\)\\
\cline{1-5}
 {0.95} &$-0.98$  &11157.72      & 1.58 &\textbf{79.53}/ 20.47\\
{0.99}& $-6.02$  & 11152.68     &0.99 &\textbf{82.40}/ 17.60 \\
 {1.03}& $-13.86$  & 11144.84     &0.78 &\textbf{84.49}/ 15.51

\\\bottomrule[1pt]\bottomrule[1pt]
 \end{tabular}
 \end{table}

The single-channel analysis of the $\Lambda_bB^{(*)}$ and $\Sigma_b^{(*)}B^{(*)}$ systems with $I(J^P)=1/2(1/2^+)$ further supports the existence of sufficient attraction in the $\Sigma_bB^*$ and $\Sigma_b^*B^*$ channels. As shown in Table \ref{tab:single1}, the loosely bound-state solutions are obtained for the single $\Sigma_bB^*$ and $\Sigma_b^*B^*$ states with $I(J^P)=1/2(1/2^+)$ when the cutoff is chosen around $\Lambda\sim1$~GeV. 

For the $\Sigma_b B^*$ state with $I(J^P)=1/2(1/2^+)$, the $^2P_{1/2}$ partial wave consistently dominates, with a probability around $90\%$, while the $^4P_{1/2}$ partial wave accounts for only $10\%$. This indicates that the spin-orbit coupling, which is indirectly mediated by the tensor force, is not particularly strong in this channel. As $\Lambda$ increases from $1.08$ GeV to $1.12$ GeV, the binding energy increases in magnitude from $-2.18$ MeV to $-8.53$ MeV, and $r_{\text{RMS}}$ shrinks from $1.22$ fm to $0.84$ fm. Meanwhile, the relative fractions of the partial-wave components remain almost unchanged.

For the $\Sigma_b^* B^*$ state with $I(J^P)=1/2(1/2^+)$, the $^2P_{1/2}$ wave also dominates, with a probability of approximately $80\%$, while $^4P_{1/2}$ contributes about $15\%$--$20\%$. By comparison, the $^4P_{1/2}$ fraction in the $\Sigma_b^* B^*$ state with $I(J^P)=1/2(1/2^+)$ is slightly larger than in the $\Sigma_b B^*$ state with $I(J^P)=1/2(1/2^+)$, because the spin-$3/2$ of the $\Sigma_b^*$ baryon and the spin-$1$ of the $B^*$ meson induce stronger tensor mixing. Nevertheless, as in the $\Sigma_b B^*$ state with $I(J^P)=1/2(1/2^+)$, the $\Sigma_b^* B^*$ state with $I(J^P)=1/2(1/2^+)$ remains clearly dominant in the $^2P_{1/2}$ partial wave.

In addition to the two loosely bound states discussed above, no loosely bound-state solutions are found for the remaining \(\Lambda_bB^{(*)}\) and \(\Sigma_b^{(*)}B^{(*)}\) systems with \(I(J^P)=1/2(1/2^+)\) in the single-channel analysis within the same cutoff region. 

\begin{figure*}[!htbp]
    \centering
    \includegraphics[width=0.85\linewidth]{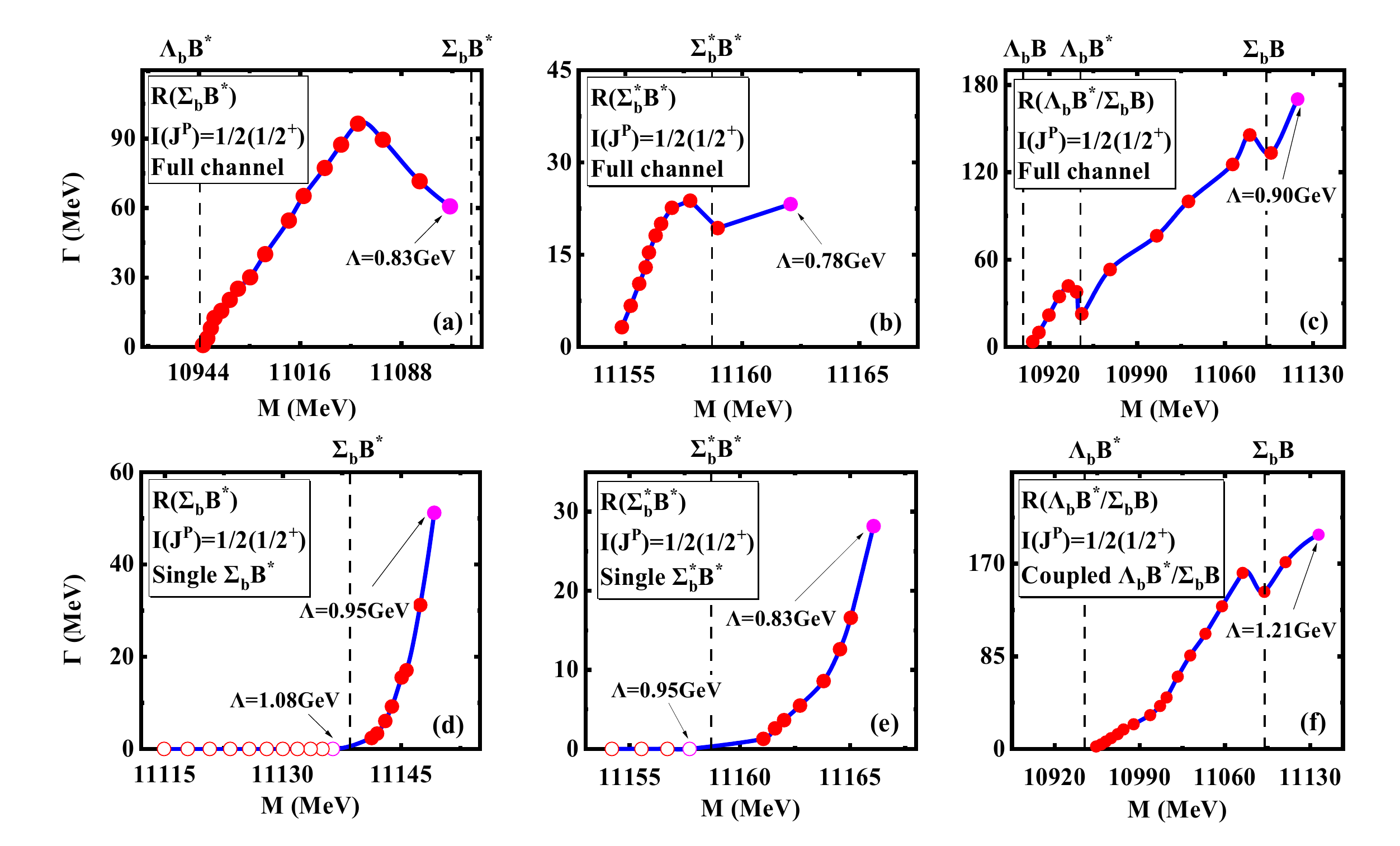}
    \caption{The cutoff dependence of the obtained resonant masses and widths from the discussed $\Lambda_bB^{(*)}/\Sigma_b^{(*)}B^{(*)}$ systems with $I(J^P)=1/2(1/2^+)$. Here, full channel represent the $\Lambda_bB^{(*)}/\Sigma_b^{(*)}B^{(*)}$ coupled system as collected in Table \ref{channel}. The single $\Sigma_b^{(*)}B^*$ and coupled $\Lambda_bB^*/\Sigma_bB$ systems correspond to the cases after sole considering the single $\Sigma_b^{(*)}B^*$ and coupled $\Lambda_bB^*/\Sigma_bB$ interactions, respectively. The pink dots correspond to the cutoff values as the resonances firstly appear. And the cutoff gap between two adjacent points is set as $\delta \Lambda=0.01$ GeV. The hollow points represent the the cutoff dependence of the obtained masses for the single $\Sigma_b^{(*)}B^*$ molecules with $I(J^P)=1/2(1/2^+)$. The dash lines label the mass thresholds for several certain channels.}
    \label{fig1}
\end{figure*}

\paragraph{\bf{Resonances and phase shifts analysis.}} We now turn to the phase-shift analysis of the $\Lambda_bB^{(*)}/\Sigma_b^{(*)}B^{(*)}$ systems with $I(J^P)=1/2(1/2^+)$, which allows us to identify possible resonance poles. Within the cutoff region around $\Lambda\sim1$ GeV, we find three hidden-bottom resonant pentaquarks. The upper panel of Fig. \ref{fig1} displays the cutoff dependence of their masses and widths. As the cutoff increases, the attraction from the OBE interaction becomes stronger, leading to a monotonic decrease of the resonance masses. In contrast, the widths do not follow a uniform trend. They develop minima near the thresholds of specific channels, indicating the importance of those channels in forming the resonances. Specifically, the $\Sigma_bB^*$ channel in panel (a), the $\Sigma_b^*B^*$ channel in panel (b), and the $\Lambda_bB^*$ and $\Sigma_bB$ channels in panel (c) play significant roles in the corresponding resonances. For convenience, we label these three states as $R(\Sigma_bB^*)[1/2(1/2^+)]$, $R(\Sigma_b^*B^*)[1/2(1/2^+)]$, and $R(\Lambda_bB^*/\Sigma_bB)[1/2(1/2^+)]$, respectively.

To further clarify the role of individual channels in forming the three resonances, we perform separate phase-shift analyses for the single $\Sigma_bB^*$ channel, the single $\Sigma_b^*B^*$ channel, and the coupled $\Lambda_bB^*/\Sigma_bB$ system. In each case, we also identify a resonance state. The lower panel of Fig.~\ref{fig1} shows the masses and widths of these resonances as functions of the cutoff (solid points). We find that the masses and widths obtained from these reduced calculations are very similar to those from the full coupled-channel analysis, which includes all $\Lambda_bB^{(*)}/\Sigma_b^{(*)}B^{(*)}$ channels. This similarity indicates that the resonances found in the single $\Sigma_bB^*$, single $\Sigma_b^*B^*$, coupled $\Lambda_bB^*/\Sigma_bB$, and full-channel calculations are the same resonance states. In addition, the cutoff values at which the resonances first appear are slightly larger in the reduced calculations than in the full-channel case. This observation suggests that the coupled-channel effects provide additional attraction and facilitate the formation of these resonance states.

The bottom panel of Fig.~\ref{fig1} also displays the masses of the single-channel \(\Sigma_bB^*\) and \(\Sigma_b^*B^*\) molecular states with \(I(J^P)=1/2(1/2^+)\) as functions of the cutoff parameter (hollow points). As the cutoff increases, the OBE attraction becomes stronger, and the resonance states gradually evolve into the loosely bound states. Consequently, the resonances \(R(\Sigma_bB^*)[1/2(1/2^+)]\) and \(R(\Sigma_b^*B^*)[1/2(1/2^+)]\) identified in our phase-shift analysis are not independent new structures. Instead, they are closely related to the corresponding single-channel bound molecular states. 

In summary, our analysis of the $\Lambda_bB^{(*)}/\Sigma_b^{(*)}B^{(*)}$ systems with $I(J^P)=1/2(1/2^+)$ yields two promising bound molecular candidates. These are the $\Sigma_b B^*$ and $\Sigma_b^* B^*$ states with $I(J^P)=1/2(1/2^+)$. In addition, we identify a resonant state denoted as $R(\Lambda_bB^*/\Sigma_bB)[1/2(1/2^+)]$.

\subsection{$I(J^P)=1/2(3/2^+)$}

\paragraph{\bf{Bound states.}} For the $\Lambda_bB^{(*)}/\Sigma_b^{(*)}B^{(*)}$ systems with $I(J^P)=1/2(3/2^+)$, the inclusion of the coupled-channel effects allows for more spin configurations due to the larger total angular momentum. Table \ref{tab:couple2} summarizes the bound-state solutions obtained for the full coupled system within a reasonable cutoff region around $\Lambda\sim1$ GeV. As the cutoff increases from $\Lambda=1.07$~GeV to $1.09$~GeV, the binding energy changes from $E=-5.50$~MeV to $-32.23$~MeV, while the RMS radius decreases from $r_{\text{RMS}}=0.49$~fm to $0.37$~fm. This behavior reflects the strengthening of the attractive interaction with increasing cutoff, consistent with the general features of the OBE potential.

\renewcommand\tabcolsep{0.22cm}
\renewcommand{\arraystretch}{1.8}
 \begin{table*}[!htbp]
\centering
\caption{The bound-state solutions for the \(\Lambda_bB^{(*)}/\Sigma_b^{(*)}B^{(*)}\) coupled system with \(I(J^P)=1/2(3/2^+)\), including the binding energy \(E\), the bound-state mass \(M\), the RMS radius \(r_{\text{RMS}}\), and the channel probabilities \(p_i\). The units of \(\Lambda\), \(E\), \(M\), and \(r_{\text{RMS}}\) are GeV, MeV, MeV, and fm, respectively.}\label{tab:couple2}
\begin{tabular}{l l l l l l l l l l}
\toprule[1pt]\toprule[1pt]
\(\Lambda\) & \(E\) &$M$   & \(r_{\text{RMS}}\) &\(\Lambda_b B(^2P_\frac{3}{2} )\)  & \(\Lambda_b B^* (^2P_\frac{3}{2}/^4P_\frac{3}{2})\)  & \(\Sigma_b B(^2P_\frac{3}{2})\) & \(\Sigma_b^* B(^4P_\frac{3}{2})\) & \(\Sigma_b B^* (^2P_\frac{3}{2}/^4P_\frac{3}{2})\) & \(\Sigma_b^* B^* (^2P_\frac{3}{2}/^4P_\frac{3}{2}/^6P_\frac{3}{2})\) \\\cline{1-10}
 1.07 & $-5.50$  & 10893.44     & 0.49&   23.33& 26.14/0.97 &4.80&0.14 & 6.44/1.02& \textbf{35.90}/1.19/0.07\\
 1.08 &$-18.14$ & 10880.80      & 0.40 &   17.70& 27.50/0.87 &4.93&0.16 & 7.98/0.93& \textbf{38.69}/1.21/0.03\\
 1.09 &$-32.23$ & 10866.71      & 0.37 &   14.94& 27.72/0.77&4.91&0.17 & 9.15/0.84& \textbf{40.30}/1.18/0.02\\
\bottomrule[1pt]\bottomrule[1pt]
\end{tabular}   
\end{table*}

The channel probabilities indicate that the bound state is predominantly composed of the $\Sigma_b^*B^*$, $\Lambda_bB^*$, and $\Sigma_bB^*$ components. Among them, the $\Sigma_b^*B^*$ channel provides the largest contribution, suggesting that this channel plays a dominant role in the formation of this bound state. The non-negligible $\Lambda_bB^*$ component further demonstrates that the coupled-channel effects are important in this system. Compared with the $I(J^P)=1/2(1/2^+)$ \(\Lambda_bB^{(*)}/\Sigma_b^{(*)}B^{(*)}\) coupled system, the higher-spin configuration in the present sector allows stronger tensor-force transitions among different partial waves. As a result, while the $^{2}P_{3/2}$ and $^{4}P_{3/2}$ partial waves remain the dominant configurations, additional mixing from other partial waves appears in the $\Sigma_b^*B^*$ channel. This indicates that the tensor force provides significant attraction and contributes importantly to the formation of this bound state.

\renewcommand\tabcolsep{0.3cm}
\renewcommand{\arraystretch}{1.8}
\begin{table}[!htbp]
\centering
\caption{The bound-state solutions for the single $\Sigma_bB^*$ and $\Sigma_b^*B^*$ states with $I(J^P)=1/2(3/2^+)$, including the binding energy $E$, the bound-state mass $M$, the RMS radius $r_{\text{RMS}}$, and the channel probabilities $p_i$. The quantities $\Lambda$, $E$, $M$, and $r_{\text{RMS}}$ are given in units of GeV, MeV, MeV, and fm, respectively.}\label{tab:single2}
\begin{tabular}{ l l l l l}
\toprule[1pt]\toprule[1pt]
 \(\Lambda\) & \(E\) &$M$ & \(r_{\text{RMS}}\) &\(\Sigma_b B^* (^2P_\frac{3}{2}/^4P_\frac{3}{2})\)   \\
\cline{1-5}
 1.11 & $-2.96$ & 11135.54        & 1.35&\textbf{54.30}/ 45.70   \\
1.13& $-5.41$ & 11133.09        &1.07 &\textbf{67.98}/ 32.02   \\
  1.15& $-8.83 $    &11129.67       &0.88 &\textbf{74.41}/ 21.59    \\\cline{1-5}
  \(\Lambda\) & \(E\) &$M$     & \(r_{\text{RMS}}\) &\(\Sigma_b^* B^* (^2P_\frac{3}{2}/^4P_\frac{3}{2}/^6P_\frac{3}{2})\)\\
\cline{1-5}
 0.96 &$-0.63 $   &  11158.07      & 1.98 &\textbf{36.13}/\textbf{63.47} / 0.40\\
 0.99& $-3.10$    &  11155.60      & 1.27&\textbf{49.36} /\textbf{49.83} / 0.81\\
1.02&$-7.03  $   & 11151.67       &0.98&\textbf{61.49}/ \textbf{37.33}/ 1.18\\\bottomrule[1pt]\bottomrule[1pt]
 \end{tabular}
 \end{table}

To further clarify the origin of the bound state, we perform the corresponding single-channel calculations. The results are listed in Table~\ref{tab:single2}. Both the single $\Sigma_bB^*$ and $\Sigma_b^*B^*$ states with $I(J^P)=1/2(3/2^+)$ yield the loosely bound-state solutions for the cutoff values around $\Lambda\sim1$~GeV. The $\Sigma_bB^*$ state with $I(J^P)=1/2(3/2^+)$ is dominated by the $^{2}P_{3/2}$ component, whereas the $\Sigma_b^*B^*$ state with $I(J^P)=1/2(3/2^+)$ exhibits stronger mixing between the $^{2}P_{3/2}$ and $^{4}P_{3/2}$ configurations. This difference arises from the larger spin of the $\Sigma_b^*$ baryon, which enhances the tensor coupling between different spin configurations. For the remaining single-channel systems, namely $\Lambda_bB^{(*)}$, $\Sigma_bB$, and $\Sigma_b^*B$ with the same quantum numbers, no loosely bound-state solutions are found, as their OBE interactions are either weakly attractive or repulsive.

\begin{figure*}[!htbp]
    \centering
    \includegraphics[width=0.85\linewidth]{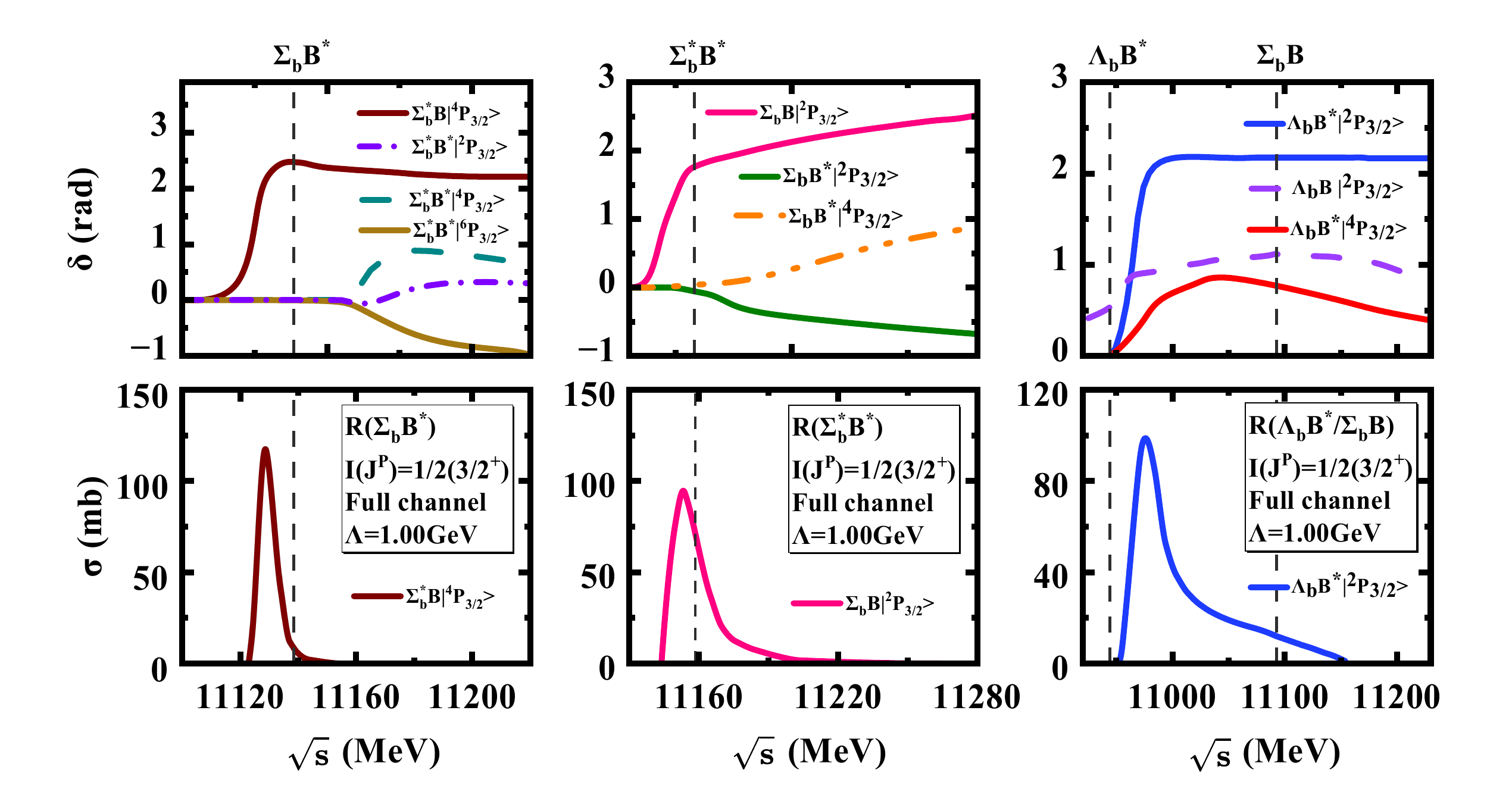}
    \caption{The phase shifts for the $\Lambda_bB^{(*)}/\Sigma_b^{(*)}B^{(*)}$ systems with $I(J^P)=1/2(3/2^+)$ as functions of the scattering energy $\sqrt{s}$.}\label{phase}
\end{figure*}

\paragraph{\bf{Resonances and phase shifts analysis.}} We further investigate the phase shifts for the $\Lambda_bB^{(*)}/\Sigma_b^{(*)}B^{(*)}$ systems with $I(J^P)=1/2(3/2^+)$. As shown in Fig.~\ref{phase}, the phase shifts for the $\Lambda_bB^*|{}^2P_{3/2}\rangle$, $\Sigma_bB|{}^2P_{3/2}\rangle$, and $\Sigma_b^*B|{}^4P_{3/2}\rangle$ channels cross $\pi/2$ within the cutoff region of about $\Lambda\sim1$~GeV. The corresponding cross sections exhibit the typical Breit-Wigner distribution of the resonance, indicating the presence of resonance structures in the $\Lambda_bB^{(*)}/\Sigma_b^{(*)}B^{(*)}$ systems with $I(J^P)=1/2(3/2^+)$. 

\begin{figure*}[!htbp]
    \centering
    \includegraphics[width=0.85\linewidth]{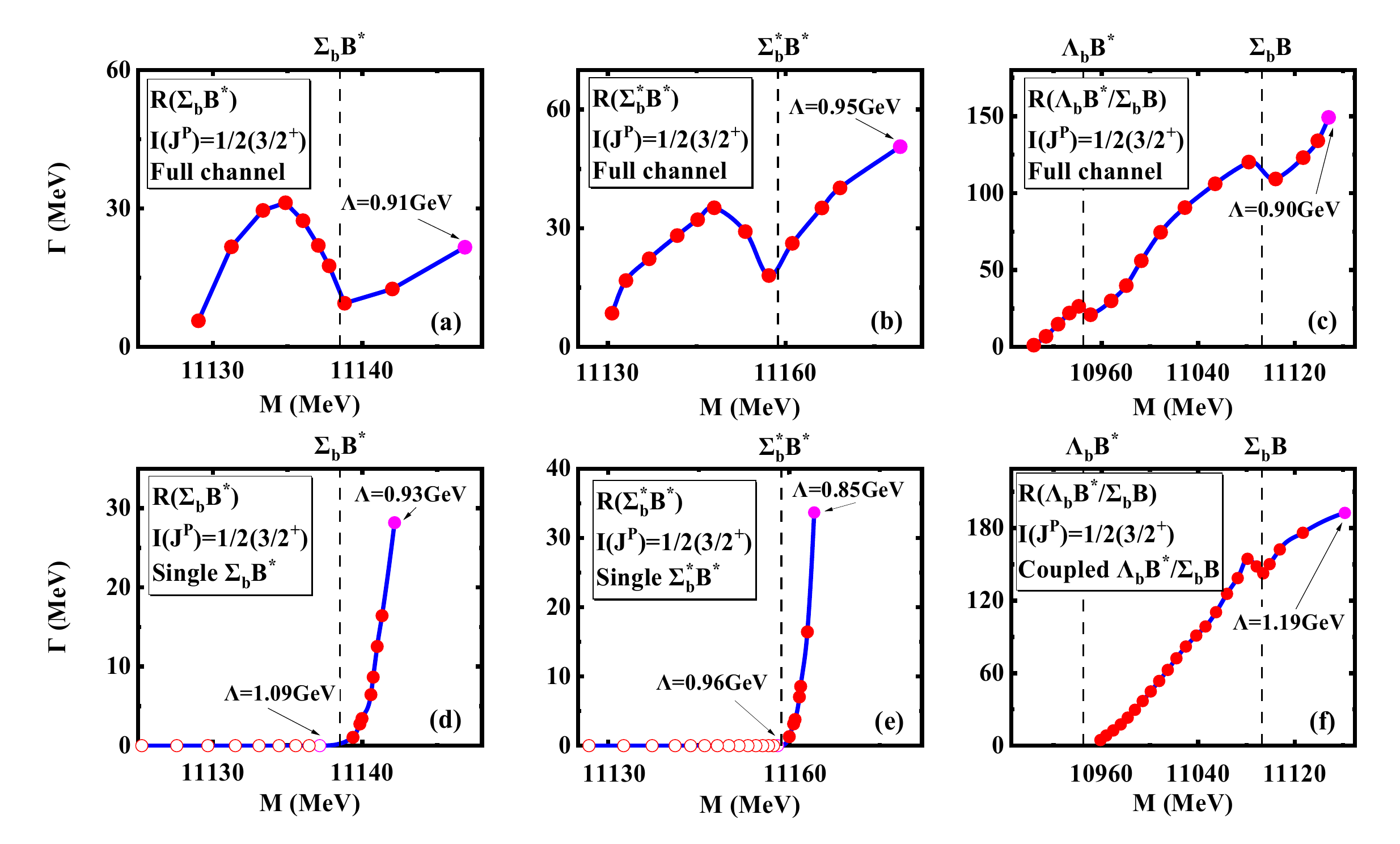}
    \caption{The cutoff dependence of the masses and widths of the resonances obtained from the $\Lambda_bB^{(*)}/\Sigma_b^{(*)}B^{(*)}$ systems with $I(J^P)=1/2(3/2^+)$. The symbols and line styles are the same as those used in Fig.~\ref{fig1}.}
    \label{fig2}
\end{figure*}

In Fig.~\ref{fig2}, we present the cutoff dependence of the masses and widths for the three resonance states. As shown in panels (a)-(c), the widths exhibit minima near the thresholds of the $\Sigma_bB^*$, $\Sigma_b^*B^*$, and $\Lambda_bB^*/\Sigma_bB$ channels, respectively. This indicates that these channels significantly influence the properties of the corresponding resonances. The phase-shift analyses performed for the single $\Sigma_bB^*$, single $\Sigma_b^*B^*$, and coupled $\Lambda_bB^*/\Sigma_bB$ channels, as displayed in panels (d)-(f), further confirm the existence of the three resonances. Their masses and widths are found to be similar to those obtained from the full coupled-channel calculations. We therefore identify these three resonances as associated with the $\Sigma_bB^*$, $\Sigma_b^*B^*$, and $\Lambda_bB^*/\Sigma_bB$ channels. Moreover, the coupled-channel interactions reduce the required cutoff values and provide additional attraction, which favors the formation of these hidden-bottom resonant pentaquarks.

In addition, as shown in Figs.~\ref{fig2}(d) and (e), we present the cutoff dependence of the resonance masses and widths, along with the molecular-state masses, in the single-channel approximation. Here, the solid and open points denote the mass and width of the resonance, and the mass of the corresponding molecular state, respectively. It is observed that as the cutoff increases, the OBE attraction becomes stronger. The masses of the two resonance states gradually decrease toward the thresholds of $\Sigma_bB^*$ and $\Sigma_b^*B^*$, respectively, while their widths also decrease. When the resonance masses approach the corresponding thresholds, the widths tend to zero, indicating that the resonances gradually evolve into the $\Sigma_bB^*$ and $\Sigma_b^*B^*$ molecular states with $I(J^P)=1/2(3/2^+)$. This further demonstrates that the resonance states $R(\Sigma_bB^*)[1/2(3/2^+)]$ and $R(\Sigma_b^*B^*)[1/2(3/2^+)]$ identified in the phase-shift analysis are not independent new structures. Instead, they are closely related to the corresponding molecular states. 

In conclusion, our analysis of both single-channel and coupled-channel dynamics predicts two promising hidden-bottom bound molecular pentaquark states with $I(J^P)=1/2(3/2^+)$, namely the $\Sigma_bB^*$ and $\Sigma_b^*B^*$ states. Additionally, one promising hidden-bottom resonant pentaquark candidate is found to arise from the $\Lambda_bB^*/\Sigma_bB$ coupling. 

\subsection{$I(J^P)=1/2(5/2^+)$}

\paragraph{\bf{Bound states.}} We next investigate the $\Lambda_bB^*/\Sigma_b^{(*)}B^{(*)}$ systems with $I(J^P)=1/2(5/2^+)$. The allowed channels in this sector include $\Lambda_bB^*$, $\Sigma_b^*B$, $\Sigma_bB^*$, and $\Sigma_b^*B^*$ configurations. The numerical results from the coupled-channel calculation are summarized in Table~\ref{tab:couple3}.

\renewcommand\tabcolsep{0.55cm}
\renewcommand{\arraystretch}{1.8}
 \begin{table*}[!htbp]
\centering
\caption{The bound-state solutions for the $\Lambda_bB^*/\Sigma_b^{(*)}B^{(*)}$ coupled system with $I(J^P)=1/2(5/2^+)$. The quantities shown include the binding energy $E$, the bound-state mass $M$, the RMS radius $r_{\text{RMS}}$, and the channel probabilities $p_i$. The cutoff parameter $\Lambda$, binding energy $E$, mass $M$, and radius $r_{\text{RMS}}$ are given in units of GeV, MeV, MeV, and fm, respectively.}\label{tab:couple3}
\begin{tabular}{l l l l l l l l }
\toprule[1pt]\toprule[1pt]
\(\Lambda\) & \(E\) & $M$& \(r_{\text{RMS}}\)  &\(\Lambda_b B^* (^4P_\frac{5}{2})\)   & \(\Sigma_b^* B(^4P_\frac{5}{2})\) & \(\Sigma_b B^* (^4P_\frac{5}{2})\) & \(\Sigma_b^* B^* (^4P_\frac{5}{2}/^6P_\frac{5}{2})\)\\\cline{1-8}
 1.10 & $-2.38$& 10942.32  & 0.69&\textbf{56.88}   &15.39    &0.41& 27.27/ 0.05\\
 1.11 & $-10.20$ &10934.50 & 0.50&\textbf{50.81}  &17.47   &0.41& 31.28/ 0.03\\
  1.12& $-19.12$& 10925.58  & 0.44  &\textbf{47.09}   &18.66   &0.40& 33.83/ 0.02\\
\bottomrule[1pt]\bottomrule[1pt]
\end{tabular}   
\end{table*}

Table \ref{tab:couple3} shows that the bound-state solutions exist for the $\Lambda_bB^*/\Sigma_b^{(*)}B^{(*)}$ coupled system with $I(J^P)=1/2(5/2^+)$ when the cutoff is varied within $\Lambda\sim 1.10$-$1.12$~GeV. As the cutoff increases, the binding energy grows in magnitude from $-2.38$~MeV to $-19.12$~MeV, and the RMS radius decreases from $0.69$~fm to $0.44$~fm. These features indicate that the resulting hadronic states are relatively compact in comparison with typical loosely bound deuteron-like systems.

The channel probability distributions indicate that the $\Lambda_bB^*$ component dominates the $\Lambda_bB^*/\Sigma_b^{(*)}B^{(*)}$ coupled system with $I(J^P)=1/2(5/2^+)$, with a fraction of approximately $47\%$-$57\%$. The $\Sigma_b^*B^*$ channel also provides a substantial contribution, particularly through the $^{4}P_{5/2}$ partial wave, whose probability gradually increases with the cutoff. In contrast, the $\Sigma_bB^*$ contribution remains relatively small. These observations suggest that the attraction in this sector originates mainly from the $\Lambda_bB^*$ and $\Sigma_b^*B^*$ interactions.

\renewcommand\tabcolsep{0.32cm}
\renewcommand{\arraystretch}{1.8}
\begin{table}[!htbp]
\centering
\caption{The bound-state solutions for the single $\Sigma_b^*B^*$ state with $I(J^P)=1/2(5/2^+)$, including the binding energy $E$, the mass $M$, the RMS radius $r_{\text{RMS}}$, and the probabilities $p_i$. The quantities $\Lambda$, $E$, $M$, and $r_{\text{RMS}}$ are given in units of GeV, MeV, MeV, and fm, respectively.}\label{tab:single3}
\begin{tabular}{ l l l l l}
\toprule[1pt]\toprule[1pt]
  \(\Lambda\) & \(E\) &$M$     & \(r_{\text{RMS}}\) &\(\Sigma_b^* B^* (^4P_\frac{5}{2}/^6P_\frac{5}{2})\)\\
\cline{1-5}
 1.10 &$-0.25$&11158.45&3.05 &1.81/ \textbf{98.19}\\
1.30&$-4.50 $&11154.20&1.49&4.49/ \textbf{95.50} \\
  1.50&$-12.51 $&11146.19&1.07&12.85/ \textbf{87.15}\\\bottomrule[1pt]\bottomrule[1pt]
 \end{tabular}
 \end{table}

To further clarify the role of individual channels, we carry out single-channel calculations for the $\Lambda_bB^*$, $\Sigma_b^*B$, $\Sigma_bB^*$, and $\Sigma_b^*B^*$ states with $I(J^P)=1/2(5/2^+)$. Among these, only the single $\Sigma_b^*B^*$ state with $I(J^P)=1/2(5/2^+)$ yields the loosely bound-state solutions within the cutoff region $\Lambda\sim1$~GeV. The corresponding bound-state solutions are listed in Table~\ref{tab:single3}. The $^{6}P_{5/2}$ partial wave constitutes the dominant component, with a probability exceeding $95\%$ in most cases. This indicates that the high-spin configuration is strongly favored in the $\Sigma_b^*B^*$ interaction. The dominant $^{6}P_{5/2}$ component arises from the spin-spin and tensor interactions between the spin-$3/2$ bottom baryon and the vector meson. For the remaining systems, namely $\Lambda_bB^*$, $\Sigma_b^*B$, and $\Sigma_bB^*$ with $I(J^P)=1/2(5/2^+)$, no loosely bound-state solutions are found within the same cutoff range.

\begin{figure}[!htbp]
    \centering
    \includegraphics[width=1.0\linewidth]{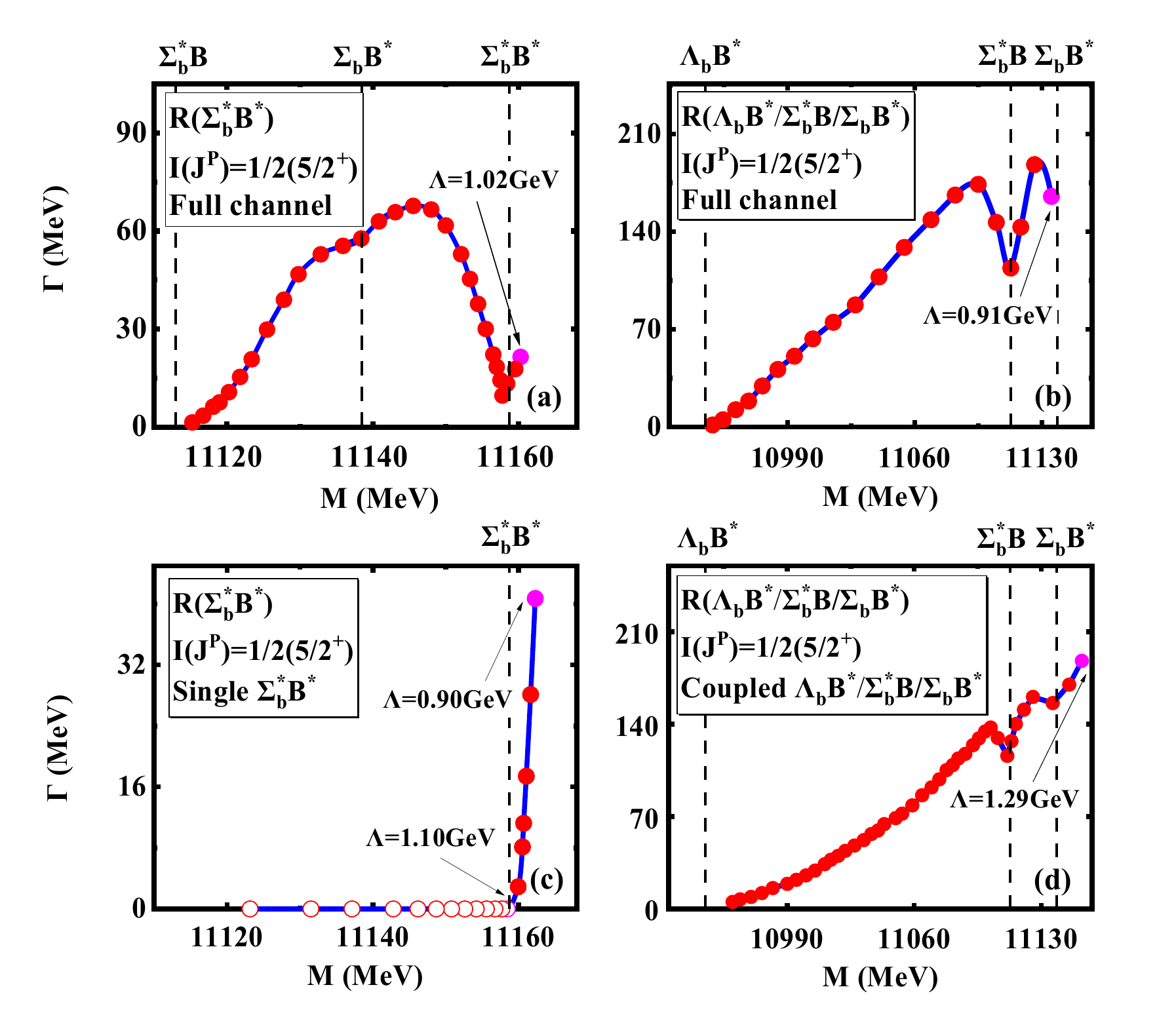}
    \caption{The cutoff dependence of the masses and widths of the resonances obtained from the $\Lambda_bB^{(*)}/\Sigma_b^{(*)}B^{(*)}$ systems with $I(J^P)=1/2(5/2^+)$. The same symbols and line styles as in Fig.~\ref{fig1} are adopted.}
    \label{fig3}
\end{figure}

\paragraph{\bf{Resonances and phase shifts analysis.}} The resonance analysis presented in Fig.~\ref{fig3} supports the existence of a hidden-bottom resonant pentaquark candidate predominantly composed of the $\Sigma_b^*B^*$ state with $I(J^P)=1/2(5/2^+)$. The resonance generated from the full coupled-channel calculation is primarily associated with the $\Sigma_b^*B^*$ interaction, and its cutoff dependence of the mass follows a trend similar to that of the single-channel result, indicating that the resonance originates mainly from the attractive $\Sigma_b^*B^*$ interaction. Moreover, as the OBE attraction becomes stronger, the resonance gradually evolves into a loosely bound state. Therefore, the resonance \(R(\Sigma_b^*B^*)[1/2(5/2^+)]\) identified in our phase-shift analysis is not an independent new structure. Instead, it is closely related to the single-channel \(\Sigma_b^*B^*\) molecular state with the same quantum numbers.

In addition to \(R(\Sigma_b^*B^*)[1/2(5/2^+)]\), we have identified another resonance from the coupled-channel phase-shift analysis for the $\Lambda_bB^{(*)}/\Sigma_b^{(*)}B^{(*)}$ systems with $I(J^P)=1/2(5/2^+)$, as shown in Figs.~\ref{fig3}(b) and~(d). We denote this resonance state as \(R(\Lambda_bB^*/\Sigma_b^*B/\Sigma_bB^*)[1/2(5/2^+)]\), because it primarily arises from the coupling among the \(\Lambda_bB^*\), \(\Sigma_b^*B\), and \(\Sigma_bB^*\) channels.

In summary, for the $\Lambda_bB^{*}/\Sigma_b^{(*)}B^{(*)}$ systems with $I(J^P)=1/2(5/2^+)$, we identify one promising hidden-bottom bound molecular pentaquark candidate and one Feshbach-type resonance. The bound state is predominantly composed of the $\Sigma_b^*B^*$ configuration, while the resonance arises primarily from the coupled-channel dynamics among the $\Lambda_bB^*$, $\Sigma_b^*B$, and $\Sigma_bB^*$ channels.

We have also examined the $\Lambda_bB^{*}/\Sigma_b^{(*)}B^{(*)}$ systems with the highest spin-parity quantum number $I(J^P)=1/2(7/2^+)$. For this sector, we scanned the cutoff parameter over a wide range from $1$ to $2$~GeV. Neither loosely bound-state solutions nor resonance poles are found from the coupled-channel Schr\"odinger equation and the phase-shift analysis. This absence indicates that the effective interactions in this high-spin channel do not provide sufficient attraction to form a molecular state or a resonant structure.

In the preceding discussions, we have systematically investigated the $\Lambda_bB^{*}/\Sigma_b^{(*)}B^{(*)}$ systems with isospin $I=1/2$ for various spin-parity assignments. Our analysis has revealed several loosely bound molecular candidates and associated resonance structures, and their properties have been discussed in detail. We now turn to the isospin-$3/2$ sector of the $\Sigma_b^{(*)}B^{(*)}$ systems. Owing to the change in the isospin factor, the OBE interactions in this sector are generally weaker, and larger cutoff values are typically required to produce the bound states or resonances. We will explore all allowed $J^P$ quantum numbers for the $I=3/2$ systems and present the corresponding bound-state and resonance solutions. The results will be compared with those from the $I=1/2$ sector to highlight the role of isospin in the formation of the hidden-bottom molecular pentaquarks.

\subsection{$I(J^P)=3/2(1/2^+)$}

\paragraph{\bf{Bound states.}} For the $\Sigma_b^{(*)}B^{(*)}$ systems with $I(J^P)=3/2(1/2^+)$, the coupled-channel results are summarized in Table~\ref{tab:couple4}. A bound state is found only when the cutoff is raised to about $\Lambda=1.21$~GeV. As the cutoff increases, the binding energy grows in magnitude from $-0.11$~MeV to $-11.12$~MeV, while the RMS radius decreases from $1.79$~fm to $0.69$~fm. The probability analysis shows that this bound state is predominantly composed of the $\Sigma_bB$ and $\Sigma_bB^*$ channels. In particular, the $^{2}P_{1/2}$ component of the $\Sigma_bB$ channel contributes more than $40\%$, and the $^{4}P_{1/2}$ component of the $\Sigma_bB^*$ channel also provides a significant contribution. In comparison with the corresponding $I=1/2$ state, the attractive interaction is considerably reduced, leading to a weaker binding effect and requiring a larger cutoff value.

\renewcommand\tabcolsep{0.5cm}
\renewcommand{\arraystretch}{1.8}
 \begin{table*}[!htbp]
\centering
\caption{The bound-state solutions for the $\Sigma_b^{(*)}B^{(*)}$ coupled system with $I(J^P)=3/2(1/2^+)$. These solutions include the binding energy $E$, the bound-state mass $M$, the RMS radius $r_{\text{RMS}}$, and the channel probabilities $p_i$. The quantities $\Lambda$, $E$, $M$, and $r_{\text{RMS}}$ are given in units of GeV, MeV, MeV, and fm, respectively.}\label{tab:couple4}
\begin{tabular}{l l l l l l l l }
\toprule[1pt]\toprule[1pt]
 \(\Lambda\) & \(E\) & $M$& \(r_{\text{RMS}}\)    & \(\Sigma_b B(^2P_\frac{1}{2})\)  & \(\Sigma_b^* B(^2P_\frac{1}{2})\) & \(\Sigma_b B^* (^2P_\frac{1}{2}/^4P_\frac{1}{2})\) & \(\Sigma_b^* B^* (^2P_\frac{1}{2}/^4P_\frac{1}{2})\) 
\\\cline{1-8}
 1.21 & $-0.11$&11092.63 & 1.79  & \textbf{61.45}  & 2.72 &1.10 / 32.98& \(\sim\)0 / 1.75\\
1.23 & $-4.61$ &11088.13& 0.86   & \textbf{47.14}  & 5.15 &1.21 /43.18& \(\sim\)0 / 3.32 \\
  1.25 & $-11.12$&11081.62 & 0.69   &{38.83}  & 7.73 &1.04 /\textbf{47.34}& \(\sim\)0 / 5.06\\
\bottomrule[1pt]\bottomrule[1pt]
\end{tabular}  
\end{table*}

\begin{figure}[!htbp]
    \centering
    \includegraphics[width=1.0\linewidth]{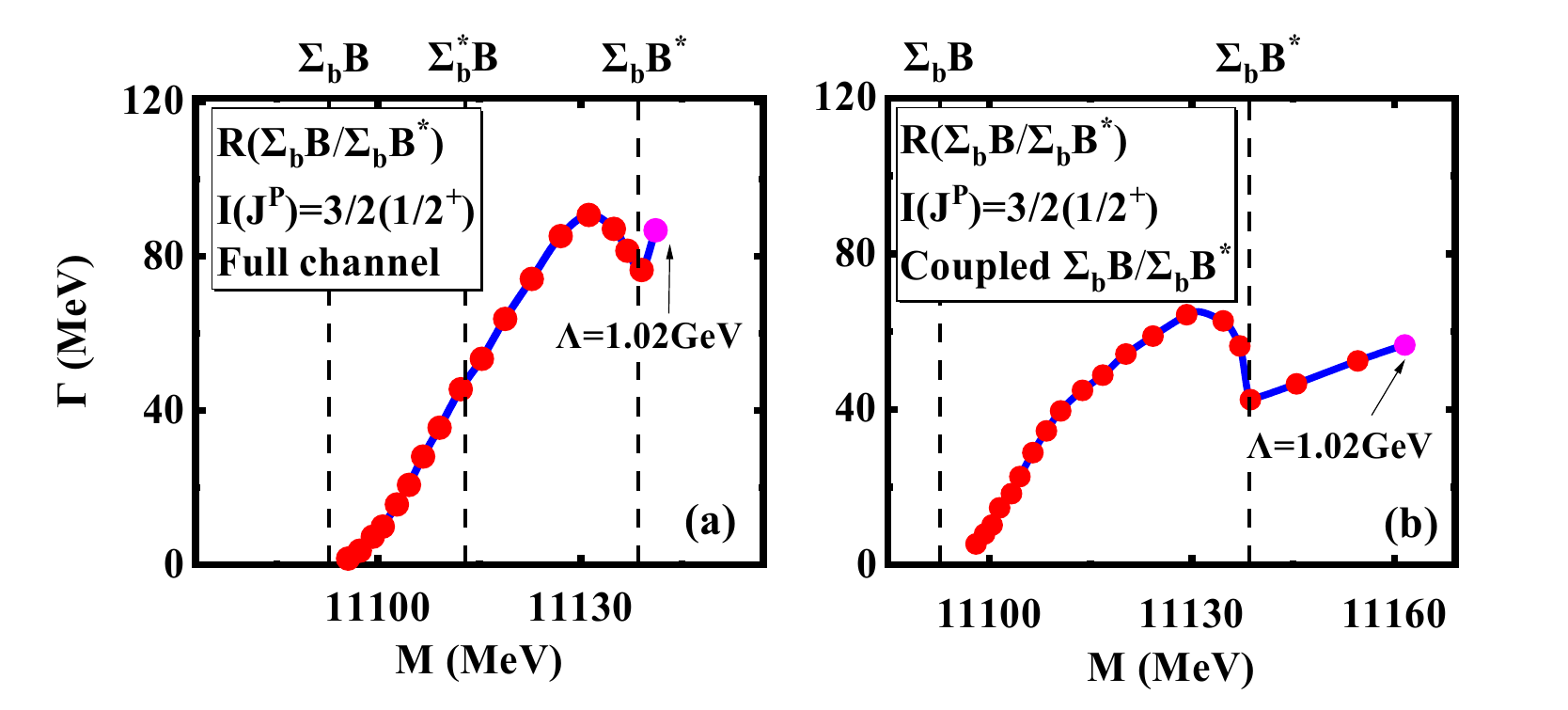}
    \caption{The cutoff dependence of the resonance masses and widths for the $\Sigma_bB^{(*)}$ coupled system with $I(J^P)=3/2(1/2^+)$. The same symbols and line styles as in Fig.~\ref{fig1} are adopted.}
    \label{fig4}
\end{figure}

\paragraph{\bf{Resonances and phase shifts analysis.}} The resonance analysis presented in Fig.~\ref{fig4} indicates that a resonant structure can emerge even when the interaction strength is relatively weak. This resonance is mainly associated with the $\Sigma_bB/\Sigma_bB^*$ coupled system with $I(J^P)=3/2(1/2^+)$. The mass and width extracted from the full coupled-channel calculation are close to those obtained from the dominant subsystems, suggesting that the resonance formation is governed primarily by specific attractive channels rather than by uniform contributions from all channels.

In summary, for the $\Sigma_b^{(*)}B^{(*)}$ systems with $I(J^P)=3/2(1/2^+)$, our analysis predicts a possible bound molecular state arising from the coupled-channel dynamics. This state corresponds to the $\Sigma_b^{(*)}B^{(*)}$ coupled system with $I(J^P)=3/2(1/2^+)$. In addition, we identify a resonant state denoted as $R(\Sigma_bB/\Sigma_bB^*)[3/2(1/2^+)]$.

\subsection{$I(J^P)=3/2(3/2^+)$}

\paragraph{\bf{Bound states.}} Table~\ref{tab:couple5} summarizes the numerical results for the $\Sigma_b^{(*)}B^{(*)}$ coupled system with $I(J^P)=3/2(3/2^+)$. Compared with the $I(J^P)=3/2(1/2^+)$ $\Sigma_b^{(*)}B^{(*)}$ coupled system, the larger total angular momentum in the present case permits more pronounced tensor mixing among the different partial-wave components.

\renewcommand\tabcolsep{0.45cm}
\renewcommand{\arraystretch}{1.8}
 \begin{table*}[!htbp]
\centering
\caption{The bound-state solutions for the $\Sigma_b^{(*)}B^{(*)}$ coupled system with $I(J^P)=3/2(3/2^+)$. The quantities shown include the binding energy $E$, the bound-state mass $M$, the RMS radius $r_{\text{RMS}}$, and the channel probabilities $p_i$. The cutoff parameter $\Lambda$, binding energy $E$, mass $M$, and radius $r_{\text{RMS}}$ are given in units of GeV, MeV, MeV, and fm, respectively.}\label{tab:couple5}
\begin{tabular}{l l l l l l l l }
\toprule[1pt]\toprule[1pt]
 \(\Lambda\) & \(E\) & $M$ & \(r_{\text{RMS}}\) & \(\Sigma_b B(^2P_\frac{3}{2})\) & \(\Sigma_b^* B(^4P_\frac{3}{2})\) & \(\Sigma_b B^* (^2P_\frac{3}{2}/^4P_\frac{3}{2})\) & \(\Sigma_b^* B^* (^2P_\frac{3}{2}/^4P_\frac{3}{2}/^6P_\frac{3}{2})\)
\\\cline{1-8}
 1.35 &$ -5.69$ &11087.05& 0.46  &7.41  & 34.14 &0.06 /  \textbf{37.96}& 0.01 / 17.86/ 2.56\\
 1.36 & $-17.16$&11075.58& 0.38  & 4.79  & 34.62 &0.03 / \textbf{39.45}& 0.01/ 19.30/ 1.80 \\
  1.37 &$ -29.79$& 11062.95 &0.34   & 3.54  & 34.67 &0.02 /\textbf{40.18}& 0.01 / 20.25/ 1.33 \\
\bottomrule[1pt]\bottomrule[1pt]
\end{tabular}   
\end{table*}

For the $\Sigma_b^{(*)}B^{(*)}$ coupled system with $I(J^P)=3/2(3/2^+)$, the bound-state solutions can be obtained for the cutoff values around $\Lambda=1.35$~GeV. As the cutoff increases, the binding energy grows in magnitude from $-5.69$~MeV to $-29.79$~MeV, while the RMS radius remains smaller than $0.5$~fm. The wave function is dominated by the $\Sigma_b^*B$ and $\Sigma_bB^*$ components, which together account for more than $70\%$ of the total probability. The sizable contribution from the $\Sigma_b^*B$ channel indicates that the spin-$3/2$ bottom baryon plays an essential role in generating sufficient attraction. In contrast, the small fraction of the $\Sigma_bB$ component suggests that the spin structure suppresses the corresponding transition. This behavior suggests that the resulting hadronic state is more compact than a typical loosely bound deuteron-like system.

The single-channel calculation for the $\Sigma_b^*B^*$ state with $I(J^P)=3/2(3/2^+)$ is presented in Table~\ref{tab:single5}, which further highlights the importance of this channel. We obtain the loosely bound-state solutions when the cutoff is raised to about $\Lambda=1.76$~GeV. This bound state is dominated by the $^{6}P_{3/2}$ partial wave, indicating that the high-spin configuration is energetically preferred. The RMS radius is also consistent with the typical size of a loosely bound state. Therefore, the $\Sigma_b^*B^*$ state with $I(J^P)=3/2(3/2^+)$ can be identified as a promising hadronic molecular candidate.

\renewcommand\tabcolsep{0.27cm}
\renewcommand{\arraystretch}{1.8}
\begin{table}[!htbp]
\centering
\caption{The bound-state solutions for the single $\Sigma_b^*B^*$ state with $I(J^P)=3/2(3/2^+)$, namely the binding energy $E$, the bound-state mass $M$, the RMS radius $r_{\text{RMS}}$, and the probabilities $p_i$ of the partial-wave components. The cutoff parameter $\Lambda$, binding energy $E$, mass $M$, and radius $r_{\text{RMS}}$ are given in units of GeV, MeV, MeV, and fm, respectively.}\label{tab:single5}
\begin{tabular}{l l l l l}
\toprule[1pt]\toprule[1pt]
 \(\Lambda\) & \(E\) & $M$& \(r_{\text{RMS}}\) &\(\Sigma_b^* B^* (^2P_\frac{3}{2}/^4P_\frac{3}{2}/^6P_\frac{3}{2})\) \\
\cline{1-5}
1.752 & $-0.56$ &11158.14& 1.20 &2.34/ 0.68/ \textbf{96.98} \\
1.760&$-6.79 $& 11151.91&0.91&1.52/ 0.28/ \textbf{98.02}  \\
  1.768&$-13.17$& 11145.53&0.78 &0.68/ 0.03/ \textbf{99.29}\\\bottomrule[1pt]\bottomrule[1pt]
 \end{tabular}
\end{table}

In addition, within the same cutoff region, no loosely bound-state solutions are found for the remaining $\Sigma_bB$, $\Sigma_b^*B$, and $\Sigma_bB^*$ systems with $I(J^P)=3/2(3/2^+)$.

\begin{figure}[!htbp]
    \centering
    \includegraphics[width=1\linewidth]{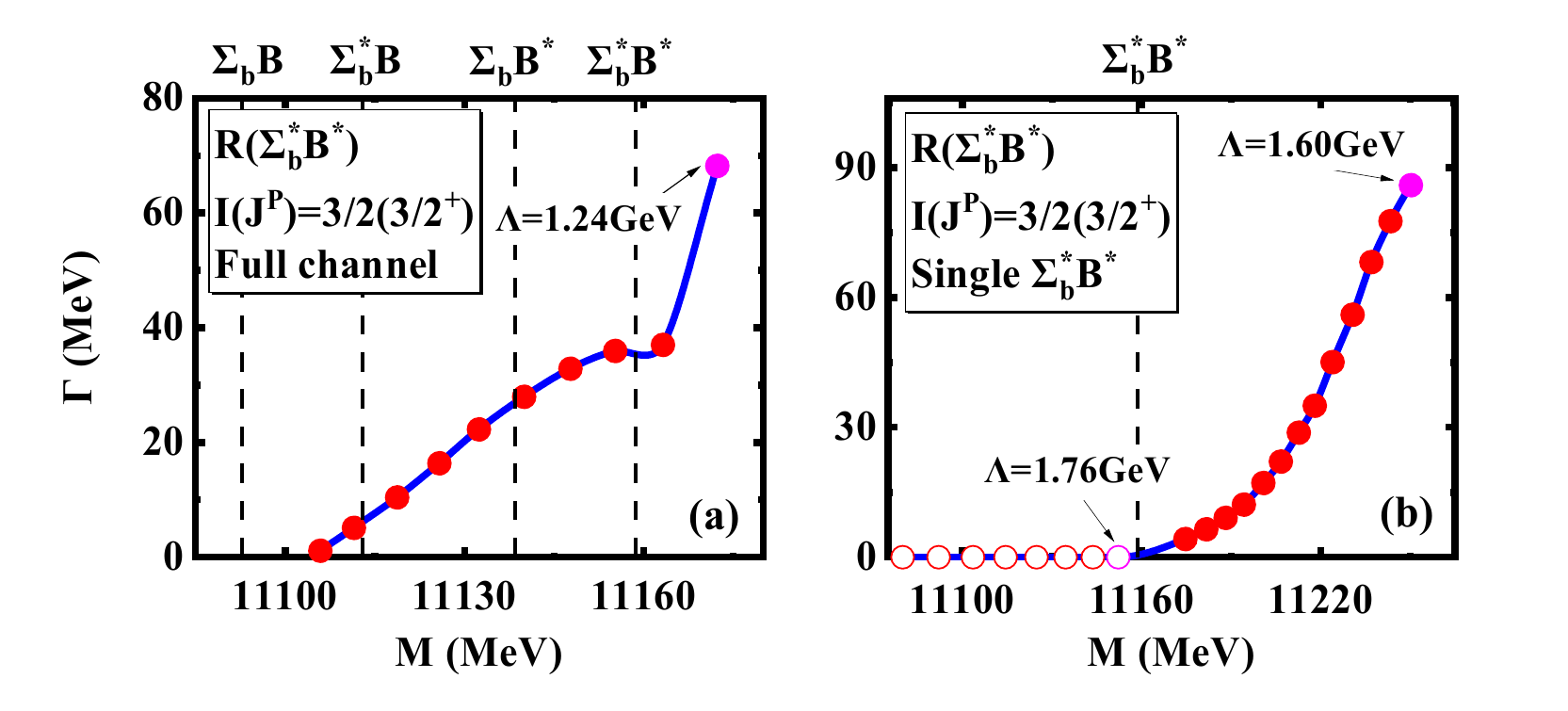}
    \caption{The cutoff dependence of the masses and widths of the resonances obtained from the $\Sigma_b^{(*)}B^{(*)}$ systems with $I(J^P)=3/2(3/2^+)$. The same symbols and line styles as in Fig.~\ref{fig1} are adopted.}
    \label{fig5}
\end{figure}

\paragraph{\bf{Resonances and phase shifts analysis.}} In Fig.~\ref{fig5}, we present the cutoff dependence of the masses and widths of the resonances obtained from the $\Sigma_b^{(*)}B^{(*)}$ system with $I(J^P)=3/2(3/2^+)$. The resonance generated from the full-channel calculation is primarily associated with the $\Sigma_b^*B^*$ interaction. The close agreement between the full-channel and single-channel results suggests that the $\Sigma_b^*B^*$ interaction provides the dominant dynamical mechanism. The larger cutoff value required in the single-channel calculation indicates that the coupled-channel effects enhance the formation of this resonance. As the cutoff parameter increases, the OBE attraction becomes stronger, and the resonance gradually evolves into the $\Sigma_b^*B^*$ molecular state with $I(J^P)=3/2(3/2^+)$. This demonstrates that the resonance $R(\Sigma_b^*B^*)[3/2(3/2^+)]$ obtained from the phase-shift analysis is not an independent new structure. Instead, it is closely related to the $\Sigma_b^*B^*$ molecular state with the same quantum numbers.

For the $\Sigma_b^{(*)}B^{(*)}$ systems with $I(J^P)=3/2(3/2^+)$, our analysis predicts only one possible hidden-bottom bound molecular pentaquark candidate. This candidate is the $\Sigma_b^*B^*$ state with $I(J^P)=3/2(3/2^+)$.
 
\subsection{$I(J^P)=3/2(5/2^+)$}

\paragraph{\bf{Bound states.}} We next investigate the $\Sigma_b^{(*)}B^{(*)}$ systems with $I(J^P)=3/2(5/2^+)$. The coupled-channel bound-state solutions are collected in Table~\ref{tab:couple6}. As in the lower-spin sectors, the isospin factor in the $I=3/2$ case reduces the effective attraction. Consequently, a larger cutoff parameter is needed to generate a bound state compared with the corresponding $I=1/2$ system.

\renewcommand\tabcolsep{0.06cm}
\renewcommand{\arraystretch}{1.8}
 \begin{table}[!htbp]
\centering
\caption{The bound-state solutions for the $\Sigma_b^{(*)}B^{(*)}$ coupled system with $I(J^P)=3/2(5/2^+)$. The quantities shown include the binding energy $E$, the bound-state mass $M$, the RMS radius $r_{\text{RMS}}$, and the channel probabilities $p_i$. The cutoff parameter $\Lambda$, binding energy $E$, mass $M$, and radius $r_{\text{RMS}}$ are given in units of GeV, MeV, MeV, and fm, respectively.}\label{tab:couple6}
\begin{tabular}{l l l l l l l  }
\toprule[1pt]\toprule[1pt]
 \(\Lambda\) & \(E\) & $M$  & \(r_{\text{RMS}}\)  & \(\Sigma_b^* B(^4P_\frac{5}{2})\) & \(\Sigma_b B^* (^4P_\frac{5}{2})\) & \(\Sigma_b^* B^* (^4P_\frac{5}{2}/^6P_\frac{5}{2})\)
\\\cline{1-7}
1.32 & $-7.12$ &11105.82 & 0.45 &\textbf{38.59}    &\textbf{38.67}& 22.73/ 0.01\\
 1.33 & $-16.24$&11096.70 & 0.40  & \textbf{37.37}   &\textbf{39.37}& 23.25/0.01 \\
  1.34 &$ -26.22$&11086.72 & 0.38  & \textbf{36.65}   &\textbf{39.76} & 23.58/ 0.01 \\
\bottomrule[1pt]\bottomrule[1pt]
\end{tabular}
\end{table}

As shown in Table~\ref{tab:couple6}, the bound-state solutions appear for the cutoff values around $\Lambda=1.32$~GeV. The binding energy grows in magnitude from $-7.12$~MeV to $-26.22$~MeV as the cutoff increases. With an RMS radius of approximately $0.4$~fm, the resulting state is relatively compact. The channel probability analysis shows that this bound state is mainly composed of the $\Sigma_b^*B$ and $\Sigma_bB^*$ components, whose probabilities are approximately comparable. This indicates that both channels provide important attractive contributions. In contrast, the $\Sigma_b^*B^*$ component contributes less than $25\%$, suggesting that this channel plays a secondary role in the formation of the bound state.

\renewcommand\tabcolsep{0.35cm}
\renewcommand{\arraystretch}{1.8}
\begin{table}[!htbp]
\centering
\caption{The bound-state solutions for the single $\Sigma_b^*B^*$ state with $I(J^P)=3/2(5/2^+)$. The quantities listed include the binding energy $E$, the bound-state mass $M$, the RMS radius $r_{\text{RMS}}$, and the partial-wave probabilities $p_i$. The cutoff parameter $\Lambda$, binding energy $E$, mass $M$, and radius $r_{\text{RMS}}$ are given in units of GeV, MeV, MeV, and fm, respectively.}\label{tab:single6}
\begin{tabular}{l l l l l}
\toprule[1pt]\toprule[1pt]
\(\Lambda\) & \(E\) & $M$ & \(r_{\text{RMS}}\) &\(\Sigma_b^* B^* (^4P_\frac{5}{2}/^6P_\frac{5}{2})\)\\
\cline{1-5}
1.68 &$-0.25$&11158.45&1.13 &2.45/ \textbf{97.55}\\
 1.69& $-6.47$ & 11152.23& 0.87&1.48/ \textbf{98.52}  \\
  1.70&$-12.45$ & 11146.25&0.70&0.17/ \textbf{99.83}\\\bottomrule[1pt]\bottomrule[1pt]
 \end{tabular}
\end{table}

The corresponding single-channel analysis for the $\Sigma_b^{(*)}B^{(*)}$ system with $I(J^P)=3/2(5/2^+)$ is presented in Table~\ref{tab:single6}. A bound state is obtained only for the single $\Sigma_b^*B^*$ state with $I(J^P)=3/2(5/2^+)$, and the required cutoff values are slightly above $1$ GeV. The wave function is almost entirely dominated by the $^{6}P_{5/2}$ partial wave, with a probability exceeding $95\%$. This clearly demonstrates that the high-spin configuration remains the most favorable structure for the $\Sigma_b^*B^*$ interaction.

\begin{figure}[!htbp]
    \centering
    \includegraphics[width=1\linewidth]{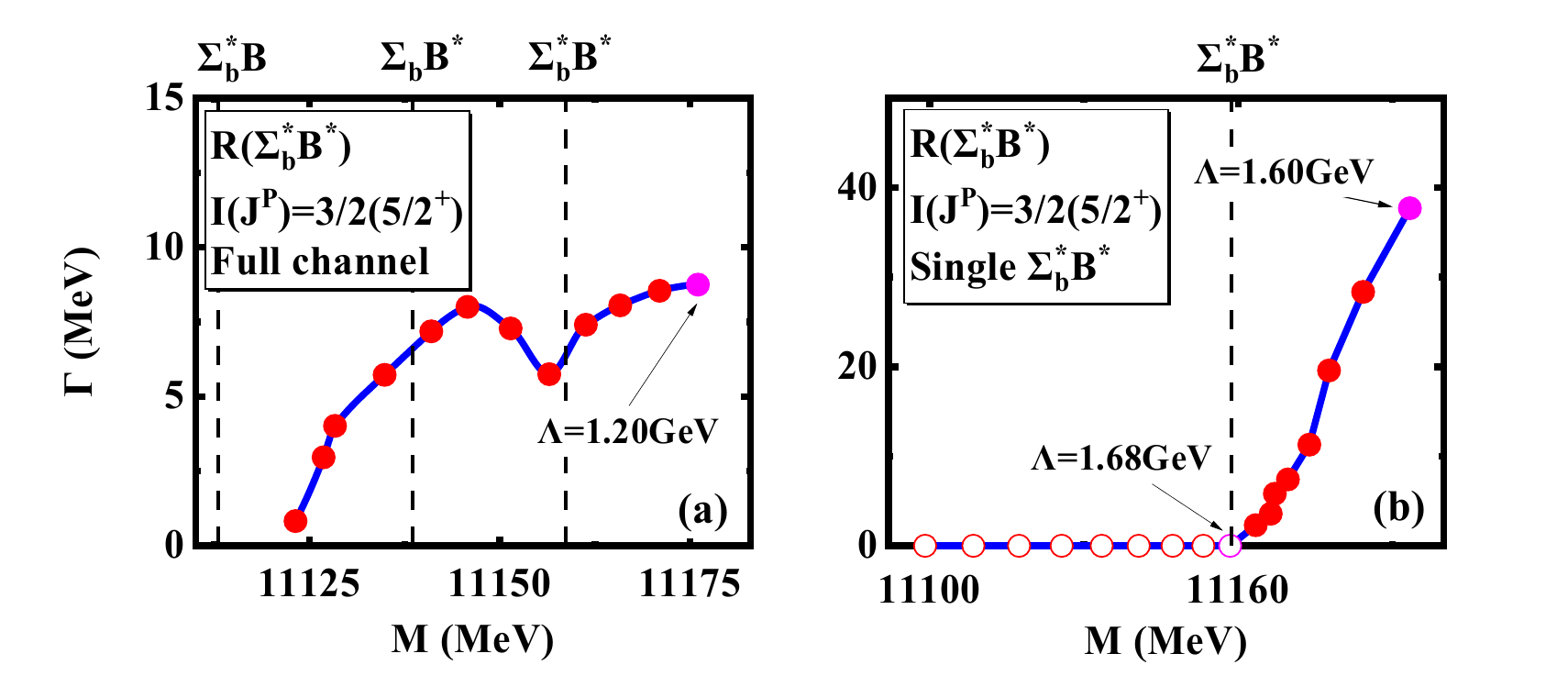}
    \caption{The cutoff dependence of the resonance masses and widths for the $\Sigma_b^{(*)}B^{(*)}$ system with $I(J^P)=3/2(5/2^+)$. The same symbols and line styles as in Fig.~\ref{fig1} are adopted.}
    \label{fig6}
\end{figure}

\paragraph{\bf{Resonances and phase shifts analysis.}} The resonance behavior shown in Fig.~\ref{fig6} indicates the existence of a possible resonant state. The resonance obtained from the full coupled-channel calculation exhibits a mass dependence similar to that of the single-channel $\Sigma_b^*B^*$ result, suggesting that this resonance is mainly generated by the $\Sigma_b^*B^*$ interaction. In addition, the coupled-channel calculation yields the resonance at a smaller cutoff, implying that the channel mixing provides extra attraction and enhances the stability of this resonance state. As the OBE attraction becomes stronger, this resonance gradually evolves into the $\Sigma_b^*B^*$ molecular state with $I(J^P)=3/2(5/2^+)$. Therefore, the resonance $R(\Sigma_b^*B^*)[3/2(5/2^+)]$ obtained from the phase-shift analysis is not an independent structure. Instead, it is intrinsically connected to the $\Sigma_b^*B^*$ molecular state with $I(J^P)=3/2(5/2^+)$.

Similar to the case of the $\Sigma_b^{(*)}B^{(*)}$ systems with $I(J^P)=3/2(3/2^+)$, our analysis yields only one possible bound molecular candidate for the $\Sigma_b^{(*)}B^{(*)}$ systems with $I(J^P)=3/2(5/2^+)$. This candidate is the $\Sigma_b^*B^*$ state with $I(J^P)=3/2(5/2^+)$.
 
\subsection{$I(J^P)=3/2(7/2^+)$}

\paragraph{\bf{Bound states.}} Finally, we consider the highest-spin configuration with $I(J^P)=3/2(7/2^+)$. In this sector, only the $\Sigma_b^*B^*$ channel is kinematically allowed, which corresponds to the $^{6}P_{7/2}$ partial wave, and the properties of this system directly reflect the intrinsic interaction strength of the $\Sigma_b^*B^*$ channel.

\renewcommand\tabcolsep{0.38cm}
\renewcommand{\arraystretch}{1.8}
\begin{table}[!htbp]
\centering
\caption{The bound-state solutions for the single $\Sigma_b^*B^*$ state with $I(J^P)=3/2(7/2^+)$. The quantities reported include the binding energy $E$, the bound-state mass $M$, the RMS radius $r_{\text{RMS}}$, and the partial-wave probabilities $p_i$. The cutoff parameter $\Lambda$, binding energy $E$, mass $M$, and radius $r_{\text{RMS}}$ are given in units of GeV, MeV, MeV, and fm, respectively.}\label{tab:single7}
\begin{tabular}{l l l l l}
\toprule[1pt]\toprule[1pt]
 \(\Lambda\) & \(E\)& $M$  & \(r_{\text{RMS}}\) &\(\Sigma_b^* B^* (^6P_\frac{7}{2})\)\\
\cline{1-5}
1.724&$-1.20$& 11157.50&0.51&\textbf{100.00}\\
1.730&$-6.28$& 11152.42&0.38&\textbf{100.00} \\
1.736&$-11.65$&11147.05&0.34 &\textbf{100.00}\\\bottomrule[1pt]\bottomrule[1pt]
 \end{tabular}
\end{table}

As shown in Table~\ref{tab:single7}, the bound-state solutions are obtained for the cutoff parameters around $\Lambda=1.72$-$1.74$~GeV. With increasing the cutoff, the binding energy grows in magnitude from $-1.20$~MeV to $-11.65$~MeV, while the RMS radius decreases from $0.51$~fm to $0.34$~fm. The relatively large cutoff required reflects the weaker attraction in the $I=3/2$ sector. Since the RMS radius remains below $0.5$~fm, the resulting bound state is relatively compact.

\begin{figure}[!htbp]
    \centering
    \includegraphics[width=0.6\linewidth]{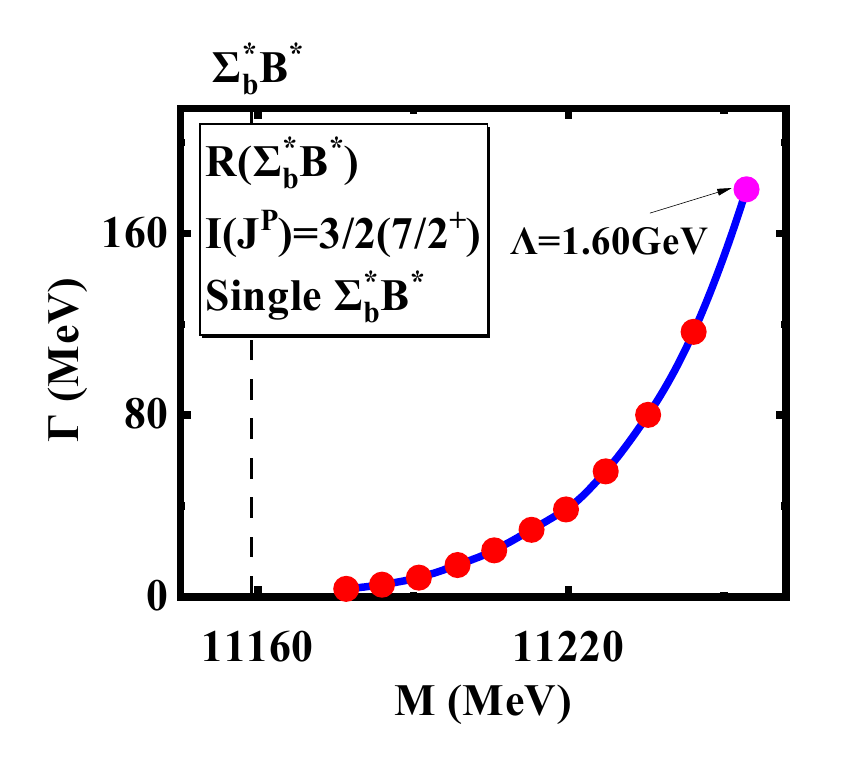}
    \caption{The cutoff dependence of the masses and widths of the resonances obtained from the $\Sigma_b^*B^*$ system with $I(J^P)=3/2(7/2^+)$. The same symbols and line styles as in Fig.~\ref{fig1} are adopted.}
    \label{fig7}
\end{figure}

\paragraph{\bf{Resonances and phase shifts analysis.}} The phase-shift analysis presented in Fig.~\ref{fig7} reveals the corresponding resonance behavior. Since only a single channel contributes, the resonance structure is fully governed by the $\Sigma_b^*B^*$ interaction. The narrow width and the pronounced variation of the phase shift indicate that this high-spin resonance state could serve as a promising target for future experimental searches.

Unlike the $\Sigma_b^{(*)}B^{(*)}$ systems with $I(J^P)=3/2(3/2^+)$ and $3/2(5/2^+)$, our analysis of the $I(J^P)=3/2(7/2^+)$ $\Sigma_b^{*}B^{*}$ system yields only one resonant candidate, denoted as $R(\Sigma_b^{*}B^{*})[3/2(7/2^+)]$.

\section{Summary}\label{sec4}

In this work, we have systematically studied the possible hidden-bottom molecular pentaquark states arising from the $P$-wave $\Lambda_bB^{(*)}/\Sigma_b^{(*)}B^{(*)}$ interactions within the OBE model. Our investigation covers both bound states and resonances. By incorporating the coupled-channel effects, we have constructed the effective potentials for all allowed quantum numbers with $J^P=1/2^+$, $3/2^+$, $5/2^+$, and $7/2^+$. We then solved the coupled-channel Schr\"odinger equations to search for the bound-state solutions and performed phase-shift analyses to identify resonance poles.

For the isospin-$1/2$ sector, we identify several promising hidden-bottom molecular pentaquark candidates within reasonable cutoff ranges. In the $J^P=1/2^+$ sector, we find two bound molecular candidates, namely the $\Sigma_b B^*$ and $\Sigma_b^* B^*$ states with $I(J^P)=1/2(1/2^+)$, and one resonant state denoted as $R(\Lambda_bB^*/\Sigma_bB)[1/2(1/2^+)]$. The coupled-channel dynamics plays an important role in reducing the required cutoff and enhancing the stability of these states. In the $J^P=3/2^+$ sector, two bound molecular candidates are found, corresponding to the $\Sigma_b B^*$ and $\Sigma_b^* B^*$ states with $I(J^P)=1/2(3/2^+)$, together with a resonant candidate arising from the $\Lambda_bB^*/\Sigma_bB$ coupling. In the $J^P=5/2^+$ sector, we obtain one bound molecular state and one Feshbach-type resonance. The former is predominantly composed of the $\Sigma_b^*B^*$ configuration, while the latter originates primarily from the coupled-channel dynamics among the $\Lambda_bB^*$, $\Sigma_b^*B$, and $\Sigma_bB^*$ channels. The wave-function analysis indicates that the $\Sigma_b^*B^*$ channel provides substantial attraction, particularly in the higher-spin configurations. The dominance of the $^{6}P_J$ partial waves in several systems highlights the crucial role of the tensor forces, in conjunction with the spin-spin interactions, in generating these high-spin molecular structures.

For the isospin-$3/2$ systems, the smaller isospin factor reduces the overall attraction and typically requires larger cutoff parameters. Despite this suppression, several bound states and resonances are still obtained, especially in channels dominated by the $\Sigma_b^*B^*$ configuration. In the $J^P=1/2^+$ sector, our analysis predicts a bound molecular state arising from coupled-channel dynamics, corresponding to the $\Sigma_b^{(*)}B^{(*)}$ coupled system with the same quantum numbers. We also identify a resonant state denoted as $R(\Sigma_bB/\Sigma_bB^*)[3/2(1/2^+)]$. In the $J^P=3/2^+$ sector, only one bound molecular pentaquark candidate is found, namely the $\Sigma_b^*B^*$ state with $I(J^P)=3/2(3/2^+)$. Similarly, in the $J^P=5/2^+$ sector, we find only one bound molecular candidate, which is the $\Sigma_b^*B^*$ state with $I(J^P)=3/2(5/2^+)$. Finally, in the $I(J^P)=3/2(7/2^+)$ sector, only one resonant candidate is identified, denoted as $R(\Sigma_b^{*}B^{*})[3/2(7/2^+)]$.

Overall, our calculations unveil a rich spectrum of positive-parity hidden-bottom molecular pentaquark states arising from the $P$-wave $\Lambda_bB^{(*)}/\Sigma_b^{(*)}B^{(*)}$ interactions. The obtained candidates exhibit diverse internal structures, spanning from the coupled-channel mixtures to nearly pure $\Sigma_b^*B^*$ configurations. These predictions provide valuable guidance for future experimental searches at LHCb and Belle II. The identification of such states would offer crucial insights into the nature of exotic hadrons and deepen our understanding of the interaction mechanisms between heavy baryons and heavy mesons.

\section*{ACKNOWLEDGMENTS}

This project is supported by the National Natural Science Foundation of China under Grant Nos. 12305139, 12305087, 12405097, 12335001, and the Xiaoxiang Scholars Programme of Hunan Normal University.

\appendix

\section{Operators in the OBE effective potentials}\label{app}

In Table \ref{potentials}, we define a serial of operators describing the spin-spin interactions and tensor forces, i.e., $\mathcal{D}_{ij}$, $\mathcal{E}_{ij}$, and $\mathcal{F}_{ij}$. The concrete expressions read as follows.
\begin{eqnarray*}
\mathcal{E}_{15} &=& \bm{\sigma} \cdot \epsilon_4^+, \quad\quad\quad\quad
\mathcal{F}_{15} = S \left( \hat{r}, \bm{\sigma}, \epsilon_4^+ \right), \quad\\
\mathcal{E}_{16} &=& \sum_{a,b}C_{\frac{1}{2},a;1,b}^{\frac{3}{2},a+b} \chi_{3, a}^{\dag} \epsilon_3^+ \cdot \epsilon_4^+ \chi_1, \quad\\
\mathcal{F}_{16} &=& \sum_{a,b}C_{\frac{1}{2},a;1,b}^{\frac{3}{2},a+b} \chi_{3, a}^{\dag} S \left( \hat{r}, \epsilon_{3,b}^+,\epsilon_4^+ \right) \chi_1.\\
\mathcal{D}_{22} &=& \chi_3^+ \chi_1 \epsilon_2 \cdot \epsilon_4^+,\\
\mathcal{E}_{23} &=& \bm{\sigma} \cdot \epsilon_2, \quad\quad\quad\quad
\mathcal{F}_{23} = S \left( \hat{r}, \bm{\sigma}, \epsilon_2\right), \quad\\
\mathcal{E}_{24} &=& \sum_{m,n} C_{\frac{1}{2},m;1,n}^{\frac{3}{2},m+n} \chi_{3,m}^+ \epsilon_{3,n}^+ \cdot \epsilon_2 \chi_1,\\
\mathcal{F}_{24} &=& \sum_{m,n} C_{\frac{1}{2},m;1,n}^{\frac{3}{2},m+n} \chi_{3,m}^+ S \left( \hat{r}, \epsilon_{3,n}^+, \epsilon_2 \right) \chi_1,\\
\mathcal{E}_{25} &=& \chi_3^+ \bm{\sigma} \cdot \left( i \epsilon_2 \times \epsilon_4^+ \right) \chi_1,\\
\mathcal{F}_{25} &=& \chi_3^{\dag} S \left( \hat{r}, \bm{\sigma}, i \bm{\epsilon}_2 \times \bm{\epsilon}_4^{\dag} \right) \chi_1,\\
\mathcal{E}_{26} &=& \sum_{m,n} C_{\frac{1}{2},m;1,n}^{\frac{3}{2},m+n} \chi_{3,m}^{\dag} \left( i \bm{\epsilon}_2 \times \bm{\epsilon}_4^{\dag} \right) \cdot \bm{\epsilon}_{3,n}^{\dag} \chi_1,\\
\mathcal{F}_{26} &=& \sum_{m,n} C_{\frac{1}{2},m;1,n}^{\frac{3}{2},m+n} \chi_{3,m}^{\dag} S \left( \hat{r}, \bm{\epsilon}_{3,n}^{\dag}, i \bm{\epsilon}_2 \times \bm{\epsilon}_4^{\dag} \right) \chi_1,\\
\mathcal{D}_{34} &=& \sum_{m,n}C_{\frac{1}{2},m;1,n}^{\frac{3}{2},m+n}
\chi_3^{m\dag}\bm{\sigma}\cdot\bm{\epsilon}_3^{n\dag}\chi_1,\\
\mathcal{E}_{35} &=& \chi_3^{\dag}\bm{\sigma}\cdot\bm{\epsilon}_4^{\dag}\chi_1,\quad\quad\quad
\mathcal{F}_{35} = \chi_3^{\dag}S(\hat{r},\bm{\sigma},\bm{\epsilon}_4^{\dag})\chi_1,\\
\mathcal{E}_{36} &=& \sum_{m,n}C_{\frac{1}{2},m;1,n}^{\frac{3}{2},m+n}\chi_3^{m\dag}
\bm{\epsilon}_4^{\dag}\cdot\left(i\bm{\sigma}\times\bm{\epsilon}_3^{n\dag}\right)\chi_1,\\
\mathcal{F}_{36} &=& \sum_{m,n}C_{\frac{1}{2},m;1,n}^{\frac{3}{2},m+n}\chi_3^{m\dag}
S\left(\hat{r},\bm{\epsilon}_4^{\dag},i\bm{\sigma}\times\bm{\epsilon}_3^{n\dag}\right)\chi_1,\\
\mathcal{D}_{44} &=& \sum_{a,b}^{m,n}C_{\frac{1}{2},a;1,b}^{\frac{3}{2},a+b}C_{\frac{1}{2},m;1,n}^{\frac{3}{2},m+n}
\chi_3^{a\dag}\bm{\epsilon}_1^n\cdot\bm{\epsilon}_3^{b\dag}\chi_1^m,\\
\mathcal{E}_{45} &=& \sum_{m,n}C_{\frac{1}{2},m;1,n}^{\frac{3}{2},m+n}\chi_3^{\dag}
\bm{\epsilon}_4^{\dag}\cdot\left(i\bm{\sigma}\times\bm{\epsilon}_1^{n}\right)\chi_1^{m},\\
\mathcal{F}_{45} &=& \sum_{m,n}C_{\frac{1}{2},m;1,n}^{\frac{3}{2},m+n}\chi_3^{\dag}
S\left(\hat{r},\bm{\epsilon}_4^{\dag},i\bm{\sigma}\times\bm{\epsilon}_1^{n}\right)\chi_1^{m},\\
\mathcal{E}_{46} &=& \sum_{a,b}^{m,n}C_{\frac{1}{2},a;1,b}^{\frac{3}{2},a+b}C_{\frac{1}{2},m;1,n}^{\frac{3}{2},m+n}
\chi_3^{a\dag}\bm{\epsilon}_4^{\dag}\cdot\left(i\bm{\epsilon}_1^{n}\times\bm{\epsilon}_3^{b\dag}\right)
\chi_1^{m},\\
\mathcal{F}_{46} &=& \sum_{a,b}^{m,n}C_{\frac{1}{2},a;1,b}^{\frac{3}{2},a+b}C_{\frac{1}{2},m;1,n}^{\frac{3}{2},m+n}
\chi_3^{a\dag}S\left(\hat{r},\bm{\epsilon}_4^{\dag},i\bm{\epsilon}_1^{n}\times\bm{\epsilon}_3^{b\dag}\right)
\chi_1^{m},\\
\mathcal{D}_{55} &=& \chi_3^{\dag}\chi_1\bm{\epsilon}_2\cdot\bm{\epsilon}_4^{\dag},\quad\quad
\mathcal{E}_{55} = \chi_3^{\dag}\bm{\sigma}\cdot\left(i\bm{\epsilon}_2\times\bm{\epsilon}_4^{\dag}\right)\chi_1,\quad\\
\mathcal{F}_{55} &=& \chi_3^{\dag}S\left(\hat{r},\bm{\sigma},i\bm{\epsilon}_2\times\bm{\epsilon}_4^{\dag}\right)\chi_1,\\
\mathcal{D}_{56} &=& \sum_{m,n}C_{\frac{1}{2},m;1,n}^{\frac{3}{2},m+n}\chi_3^{m\dag}
\left(\bm{\sigma}\cdot\bm{\epsilon}_3^{n\dag}\right)
\left(\bm{\epsilon}_2\cdot\bm{\epsilon}_4\right)\chi_1,\\
\mathcal{E}_{56} &=& \sum_{m,n}C_{\frac{1}{2},m;1,n}^{\frac{3}{2},m+n}\chi_3^{m\dag}
\left(\bm{\sigma}\times\bm{\epsilon}_3^{n\dag}\right)
\cdot\left(\bm{\epsilon}_2\times\bm{\epsilon}_4\right)\chi_1,\\
\mathcal{F}_{56} &=& \sum_{m,n}C_{\frac{1}{2},m;1,n}^{\frac{3}{2},m+n}\chi_3^{m\dag}
S\left(\hat{r},\bm{\sigma}\times\bm{\epsilon}_3^{n\dag},
\bm{\epsilon}_2\times\bm{\epsilon}_4\right)\chi_1,\\
\mathcal{D}_{66} &=& \sum_{a,b}^{m,n}C_{\frac{1}{2},a;1,b}^{\frac{3}{2},a+b}C_{\frac{1}{2},m;1,n}^{\frac{3}{2},m+n}
\chi_3^{a\dag}\left(\bm{\epsilon}_1^n\cdot\bm{\epsilon}_3^{b\dag}\right)
\left(\bm{\epsilon}_2\cdot\bm{\epsilon}_4\right)\chi_1^m,\\
\mathcal{E}_{66} &=& \sum_{a,b}^{m,n}C_{\frac{1}{2},a;1,b}^{\frac{3}{2},a+b}C_{\frac{1}{2},m;1,n}^{\frac{3}{2},m+n}
\chi_3^{a\dag}\left(\bm{\epsilon}_1^n\times\bm{\epsilon}_3^{b\dag}\right)
\cdot\left(\bm{\epsilon}_2\times\bm{\epsilon}_4\right)\chi_1^m,\nonumber\\\\
\mathcal{F}_{66} &=& \sum_{a,b}^{m,n}C_{\frac{1}{2},a;1,b}^{\frac{3}{2},a+b}C_{\frac{1}{2},m;1,n}^{\frac{3}{2},m+n}
\chi_3^{a\dag}S\left(\hat{r},\bm{\epsilon}_1^n\times\bm{\epsilon}_3^{b\dag},
\bm{\epsilon}_2\times\bm{\epsilon}_4\right)\chi_1^m.\nonumber\\
\end{eqnarray*}

When we perform numerical calculations, the above operators will be replaced by matrices elements, as summarized in Table \ref{elements1}.

\begin{table*}[!htbp]
\centering
\renewcommand\tabcolsep{0.15cm}
\renewcommand{\arraystretch}{1.7}
\caption{ Matrix elements $\langle f | \mathcal{O} | i \rangle$ for all the operators $\mathcal{O}$.} \label{elements1}
\begin{tabular}{ccccccc}
\toprule[1pt]\toprule[1pt]
 $J^P$& $\mathcal{E}_{15}$
 & $\mathcal{F}_{15}$
 & $\mathcal{E}_{16}$
 & $\mathcal{F}_{16}$
 & $\mathcal{E}_{23}$
 & $\mathcal{F}_{23}$ \\
\hline
$1/2^+$
& $\begin{pmatrix} \sqrt{3} & 0 \end{pmatrix}$
& $\begin{pmatrix} 0 & -\sqrt{6} \end{pmatrix}$
& $\begin{pmatrix} -\sqrt{2} & 0 \end{pmatrix}$
& $\begin{pmatrix} 0 & -\sqrt{\dfrac{2}{5}} \end{pmatrix}$
& $\begin{pmatrix} \sqrt{3} \\ 0 \end{pmatrix}$
& $\begin{pmatrix} 0 \\ -\sqrt{6} \end{pmatrix}$ \\
$3/2^+$
& $\begin{pmatrix} \sqrt{3} & 0 \end{pmatrix}$
& $\begin{pmatrix} 0 & \sqrt{\dfrac{3}{5}} \end{pmatrix}$
& $\begin{pmatrix} -\sqrt{2} & 0 & 0 \end{pmatrix}$
& $\begin{pmatrix} 0 & \dfrac{1}{5} & -\dfrac{3\sqrt{6}}{5} \end{pmatrix}$
& $\begin{pmatrix} \sqrt{3} \\ 0 \end{pmatrix}$
& $\begin{pmatrix} 0 \\ \sqrt{\dfrac{3}{5}} \end{pmatrix}$\\\hline
$J^P$& $\mathcal{E}_{24}$
& $\mathcal{F}_{24}$
& $\mathcal{E}_{25}$
& $\mathcal{F}_{25}$
& $\mathcal{E}_{26}$
& $\mathcal{F}_{26}$

 \\
\hline
$1/2^+$
& $\begin{pmatrix} 0 \\ 1 \end{pmatrix}$
& $\begin{pmatrix} 0 \\ \dfrac{1}{5} \end{pmatrix}$
& $\begin{pmatrix} -2 & 0 \\ 0 & 1 \end{pmatrix}$
& $\begin{pmatrix} 0 & -\sqrt{2} \\ -\sqrt{2} & -2 \end{pmatrix}$
& $\begin{pmatrix} -\sqrt{\dfrac{2}{3}} & 0 \\ 0 & -\sqrt{\dfrac{5}{3}} \end{pmatrix}$
& $\begin{pmatrix} 0 & -4\sqrt{\dfrac{2}{15}} \\ -\dfrac{1}{\sqrt{3}} & \dfrac{1}{\sqrt{15}} \end{pmatrix}$
\\
$3/2^+$
& $\begin{pmatrix} 0 \\ 1 \end{pmatrix}$
& $\begin{pmatrix} \dfrac{2}{5\sqrt{5}} \\ -\dfrac{8}{25} \end{pmatrix}$
& $\begin{pmatrix} -2 & 0 \\ 0 & 1 \end{pmatrix}$
& $\begin{pmatrix} 0 & \dfrac{1}{\sqrt{5}} \\ \dfrac{1}{\sqrt{5}} & \dfrac{8}{5} \end{pmatrix}$
& $\begin{pmatrix} -\sqrt{\dfrac{2}{3}} & 0 & 0 \\ 0 & -\sqrt{\dfrac{5}{3}} & 0 \end{pmatrix}$
& $\begin{pmatrix} 0 & \dfrac{4}{5\sqrt{3}} & \dfrac{3\sqrt{2}}{5} \\ \dfrac{1}{\sqrt{30}} & -\dfrac{4}{5\sqrt{15}} & -\dfrac{21}{5\sqrt{10}} \end{pmatrix}$
\\

$5/2^+$
& $\begin{pmatrix} 1 \end{pmatrix}$
& $\begin{pmatrix} \dfrac{1}{5} \end{pmatrix}$
& $\begin{pmatrix} 1 \end{pmatrix}$
& $\begin{pmatrix} -\dfrac{2}{5} \end{pmatrix}$
& $\begin{pmatrix} -\sqrt{\dfrac{5}{3}} & 0 \end{pmatrix}$
& $\begin{pmatrix} \dfrac{1}{5\sqrt{15}} & \dfrac{3\sqrt{5}}{5} \end{pmatrix}$\\
\hline
$J^P$& $\mathcal{E}_{35}$
& $\mathcal{F}_{35}$
& $\mathcal{E}_{36}$
& $\mathcal{F}_{36}$
& $\mathcal{D}_{44}$
& $\mathcal{E}_{45}$
 \\
\hline
$1/2^+$
& $\begin{pmatrix} \sqrt{3} \\ 0 \end{pmatrix}$
& $\begin{pmatrix} 0 \\ -\sqrt{6} \end{pmatrix}$
& $\begin{pmatrix} \sqrt{2} \\ 0 \end{pmatrix}$
& $\begin{pmatrix} 0 \\ \sqrt{\dfrac{2}{5}} \end{pmatrix}$
& $\begin{pmatrix} 1 \end{pmatrix}$
& $\begin{pmatrix} 0 & 1 \end{pmatrix}$\\
$3/2^+$
& $\begin{pmatrix} \sqrt{3} \\ 0 \end{pmatrix}$
& $\begin{pmatrix} 0 \\ \sqrt{\dfrac{3}{5}}  \end{pmatrix}$
& $\begin{pmatrix} \sqrt{2} \\ 0 \\0 \end{pmatrix}$
& $\begin{pmatrix} 0 \\ -\dfrac{1}{5} \\ \dfrac{3\sqrt{6}}{5} \end{pmatrix}$
& $\begin{pmatrix} 1 \end{pmatrix}$
& $\begin{pmatrix} 0 & 1 \end{pmatrix}$
\\

$5/2^+$
& \ldots
& \ldots
& \ldots
& \ldots
& $\begin{pmatrix} 1 \end{pmatrix}$
& $\begin{pmatrix} 1 \end{pmatrix}$\\
\hline
$J^P$& $\mathcal{F}_{45}$
& $\mathcal{E}_{46}$
& $\mathcal{F}_{46}$
& $\mathcal{D}_{55}$
& $\mathcal{E}_{55}$
& $\mathcal{F}_{55}$
 \\\hline
$1/2^+$
& $\begin{pmatrix} -\sqrt{2} \end{pmatrix}$& $\begin{pmatrix} 0 & \sqrt{\dfrac{5}{3}} \end{pmatrix}$
& $\begin{pmatrix} \dfrac{1}{\sqrt{3}} & -\dfrac{4}{\sqrt{15}} \end{pmatrix}$
& $\begin{pmatrix} 1 & 0 \\ 0 & 1 \end{pmatrix}$
& $\begin{pmatrix} -2 & 0 \\ 0 & 1 \end{pmatrix}$
& $\begin{pmatrix} 0 & -\sqrt{2} \\ -\sqrt{2} & -2 \end{pmatrix}$\\
$3/2^+$
& $\begin{pmatrix} \dfrac{1}{\sqrt{5}} & -\dfrac{4}{5} \end{pmatrix}$
& $\begin{pmatrix} 0 & \sqrt{\dfrac{5}{3}} & 0 \end{pmatrix}$
& $\begin{pmatrix} -\dfrac{1}{\sqrt{30}} & \dfrac{16}{5\sqrt{15}} & -\dfrac{21}{5\sqrt{10}} \end{pmatrix}$ 
& $\begin{pmatrix} 1 & 0 \\ 0 & 1 \end{pmatrix}$
& $\begin{pmatrix} -2 & 0 \\ 0 & 1 \end{pmatrix}$
& $\begin{pmatrix} 0 & \dfrac{1}{\sqrt{5}} \\ \dfrac{1}{\sqrt{5}} & \dfrac{8}{5} \end{pmatrix}$
\\
$5/2^+$
& $\begin{pmatrix} \dfrac{1}{5} \end{pmatrix}$
& $\begin{pmatrix} \sqrt{\dfrac{5}{3}} & 0 \end{pmatrix}$
& $\begin{pmatrix} -\dfrac{4}{5\sqrt{15}} & \dfrac{3\sqrt{5}}{5} \end{pmatrix}$
& $\begin{pmatrix} 1  \end{pmatrix}$
& $\begin{pmatrix} 1 \end{pmatrix}$
& $\begin{pmatrix} -\dfrac{2}{5} \end{pmatrix}$\\
\hline

$J^P$&\ldots  & $\mathcal{E}_{56}$
& $\mathcal{F}_{56}$
& $\mathcal{D}_{66}$
& $\mathcal{E}_{66}$
& $\mathcal{F}_{66}$ \\\hline
$1/2^+$
&\ldots& $\begin{pmatrix} \sqrt{\dfrac{2}{3}} & 0 \\ 0 & \sqrt{\dfrac{5}{3}} \end{pmatrix}$
& $\begin{pmatrix} 0 & 4\sqrt{\dfrac{2}{15}} \\ \dfrac{1}{\sqrt{3}} & -\dfrac{1}{\sqrt{15}} \end{pmatrix}$
& $\begin{pmatrix} 1 & 0 \\ 0 & 1 \end{pmatrix}$
& $\begin{pmatrix} \dfrac{5}{3} & 0 \\ 0 & \dfrac{2}{3} \end{pmatrix}$
& $\begin{pmatrix} 0 & -\dfrac{7}{3\sqrt{5}} \\ -\dfrac{7}{3\sqrt{5}} & \dfrac{16}{15} \end{pmatrix}$\\
$3/2^+$
&\ldots& $\begin{pmatrix} \sqrt{\dfrac{2}{3}} & 0 & 0 \\ 0 & \sqrt{\dfrac{5}{3}} & 0 \end{pmatrix}$
& $\begin{pmatrix} 0 & -\dfrac{4}{5\sqrt{3}} & -\dfrac{3\sqrt{2}}{5} \\ -\dfrac{1}{\sqrt{30}} & \dfrac{4}{5\sqrt{15}} & \dfrac{21}{5\sqrt{10}} \end{pmatrix}$
& $\begin{pmatrix} 1 & 0 & 0 \\ 0 & 1 & 0 \\ 0 & 0 & 1 \end{pmatrix}$
& $\begin{pmatrix} \dfrac{5}{3} & 0 & 0 \\ 0 & \dfrac{2}{3} & 0 \\ 0 & 0 & -1 \end{pmatrix}$
& $\begin{pmatrix} 0 & \dfrac{7}{15\sqrt{2}} & -\dfrac{2\sqrt{3}}{5} \\ \dfrac{7}{15\sqrt{2}} & -\dfrac{64}{75} & -\dfrac{7\sqrt{6}}{25} \\ -\dfrac{2\sqrt{3}}{5} & -\dfrac{7\sqrt{6}}{25} & \dfrac{28}{25} \end{pmatrix}$ \\
$5/2^+$
&\ldots& $\begin{pmatrix} \sqrt{\dfrac{5}{3}} & 0 \end{pmatrix}$
& $\begin{pmatrix} -\dfrac{1}{5\sqrt{15}} & -\dfrac{3\sqrt{5}}{5} \end{pmatrix}$
& $\begin{pmatrix} 1 & 0 \\ 0 & 1 \end{pmatrix}$
& $\begin{pmatrix} \dfrac{2}{3} & 0 \\ 0 & -1 \end{pmatrix}$
& $\begin{pmatrix} \dfrac{16}{75} & \dfrac{\sqrt{21}}{25} \\ \dfrac{\sqrt{21}}{25} & -\dfrac{32}{25} \end{pmatrix}$ \\
$7/2^+$
&\ldots& \ldots
 & \ldots
& $\begin{pmatrix} 1 \end{pmatrix}$
& $\begin{pmatrix} -1 \end{pmatrix}$
& $\begin{pmatrix} \dfrac{2}{5} \end{pmatrix}$ \\
\bottomrule[1pt]\bottomrule[1pt]
\end{tabular}
\end{table*}

\end{document}